\documentclass[12pt,a4paper]{article}
\usepackage{graphicx}
\usepackage[T1]{fontenc}
\usepackage[utf8]{inputenc}
\usepackage{textcomp}
\usepackage[sc,osf]{mathpazo}
\usepackage{a4wide}  
\usepackage{latexsym,amsthm,amsfonts,amsmath,mathrsfs,amssymb}
\usepackage{dsfont}
\usepackage{accents}
\usepackage[nosort]{cite}
\usepackage{booktabs} 
\usepackage[unicode,implicit]{hyperref}
\hypersetup{%
  pdftitle    = {Black hole chemistry, the cosmological constant and the embedding tensor}
  pdfkeywords = {Black holes, thermodynamics, Noether charge, Wald entropy,
    first law, black hole chemistry, embedding tensor, duality},
  pdfauthor   = {Patrick Meessen, Dimitrios Mitsios and Tom\'as Ort\'{\i}n},
  plainpages  = true,
  colorlinks  = true,
  citecolor   = blue,
  urlcolor    = red,
  linkcolor   = black
}
\newcommand{\hepth}[1]{{\tt
\href{http://www.arXiv.org/abs/hep-th/#1}{hep-th/#1}}}
\newcommand{\grqc}[1]{{\tt
\href{http://www.arXiv.org/abs/gr-qc/#1}{gr-qc/#1}}}

\newcommand{\arxiv}[1]{{\tt arXiv:\href{http://www.arXiv.org/abs/#1}{#1}}}
\newcommand{\doi}[1]{{\tt DOI:\href{http://dx.doi.org/#1}{#1}}}

\makeatletter
\@addtoreset{equation}{section}
\makeatother

\begin{document}

\begin{flushright}
\small
IFT-UAM/CSIC-22-003\\
March 25\textsuperscript{th}, 2022\\
\normalsize
\end{flushright}

\vspace{1cm}

\begin{center}

  {\Large {\bf Black hole chemistry, the cosmological constant\\[.3cm]
      and the embedding tensor}}

\vspace{1.5cm}

\renewcommand{\thefootnote}{\alph{footnote}}

{\sl\large Patrick Meessen,$^{1,2,}$}\footnote{Email: {\tt meessenpatrick[at]uniovi.es}}
{\sl\large Dimitrios Mitsios}$^{3,4}$\footnote{Email: {\tt  dimitrios.mitsios[at]ipht.fr}}
{\sl\large  and Tom\'{a}s Ort\'{\i}n,}$^{3,}$\footnote{Email: {\tt tomas.ortin[at]csic.es}}

\setcounter{footnote}{0}
\renewcommand{\thefootnote}{\arabic{footnote}}
\vspace{1cm}

${}^{1}${\it HEP Theory Group, Departamento de F\'{\i}sica, Universidad de Oviedo\\
  Avda.~Calvo Sotelo s/n, E-33007 Oviedo, Spain}\\

\vspace{0.2cm}

${}^{2}${\it Instituto Universitario de Ciencias y Tecnolog\'{\i}as Espaciales
  de Asturias (ICTEA)\\ Calle de la Independencia, 13, E-33004 Oviedo, Spain}\\

\vspace{0.2cm}

${}^{3}${\it Instituto de F\'{\i}sica Te\'orica UAM/CSIC\\
C/ Nicol\'as Cabrera, 13--15,  C.U.~Cantoblanco, E-28049 Madrid, Spain}

\vspace{0.2cm}
${}^4${\it  Universit\'e Paris-Saclay, CNRS, CEA, Institut de Physique Th\'{e}orique,\\ 91191, Gif-sur-Yvette, France}

\vspace{1cm}


{\bf Abstract}
\end{center}
\begin{quotation}
  {\small We study black-hole thermodynamics in theories that contain
    dimensionful constants such as the cosmological constant or coupling
    constants in Wald's formalism. The most natural way to deal with these
    constants is to promote them to scalar fields introducing a $(d-1)$-form
    Lagrange multiplier that forces them to be constant on-shell. These
    $(d-1)$-form potentials provide a dual description of them and, in the
    context of superstring/supergravity theories, a higher-dimensional
    origin/explanation. In the context of gauged supergravity theories, all
    these constants can be collected in the \textit{embedding tensor}. We
    show in an explicit 4-dimensional example that the embedding tensor can
    also be understood as a thermodynamical variable that occurs in the Smarr
    formula in a duality-invariant fashion. This establishes an interesting
    link between black-hole thermodynamics, gaugings and compactifications in
    the context of superstring/supergravity theories.}
\end{quotation}

\newpage
\pagestyle{plain}

\tableofcontents


\section*{Introduction}

The realization in Refs.~\cite{Kastor:2008xb,Kastor:2009wy} that the
cosmological constant can be considered as a thermodynamical variable in the
context of black-hole physics has lead to a host of new developments
encompassed in the field of \textit{black-hole chemistry}.\footnote{For
  reviews with many references see
  Refs.~\cite{Mann:2015luq,Kubiznak:2016qmn}.} As shown in
\cite{Kastor:2010gq}, other constants defining a theory can also be
seen as thermodynamical variables; in said reference, these constants
occur as coefficients of higher-order curvature terms that are
Lovelock densities.
\par
Clearly, the same ideas can be applied to $f(R)$ theories. Such
theories can, however, be
rewritten as theories of gravity coupled to a real scalar field with a non-trivial
scalar potential. The form of the potential is related to the function $f(R)$ and
therefore contains the same constants as the function $f(R)$,
In Ref.~\cite{Ortin:2021ade}, one of the authors proposed that the
constants occurring in general scalar
potentials can also be seen as thermodynamical variables in black-hole
physics. In gauged supergravity,\footnote{For a review with many references,
  see Ref.~\cite{Trigiante:2016mnt}.} though, these constants are related to
the gauge coupling constants, which can be generically represented by the
so-called \textit{embedding tensor}.\footnote{For a pedagogical introduction
  and references, see, for instance, Ref.~\cite{Ortin:2015hya}.} This lead
us to conjecture that the embedding tensor itself should also be regarded as a
thermodynamical variable. Testing this conjecture is one of the main goals of
this paper.

Wald's formalism \cite{Lee:1990nz,Wald:1993nt,Iyer:1994ys} provides an
efficient method to study black-hole thermodynamics, once the gauge freedoms
of the fields have correctly been taken into account as explained in
Refs.~\cite{Elgood:2020svt,Elgood:2020mdx,Elgood:2020nls}.\footnote{A
  different approach to handle gauge charges based on a ``solution phase
  space'' has been proposed in Ref.~\cite{Hajian:2015xlp}.} It is only by
doing this that one obtains the work terms in the first law of black-hole
mechanics. However, one does not get all the work terms that are usually
admitted in the literature\footnote{See,
  \textit{e.g.}~Ref.~\cite{Gibbons:1996af}, in which terms proportional to the
  variations of the moduli are included. This inclusion has been contested in
  Ref.~\cite{Hajian:2016iyp}.}  because there are no gauge symmetries
associated to all of them: the gauge symmetry of the electromagnetic field
gives rise to a work term of the form $\Phi\delta Q$, where $Q$ is the
electric charge and $\Phi$ is the electric potential on the event horizon, but
there is no additional gauge symmetry to give rise to the dual term
$\tilde{\Phi}\delta P$ where $P$ is the magnetic charge and $\tilde{\Phi}$ is
the magnetic potential on the horizon. There is no term proportional to the
variation of the moduli, either, because there are no gauge symmetries
associated to them. In addition, closer to our concerns in this paper, there
is no term involving variations of the cosmological constant, for the same
reason.\footnote{ One may argue that, perhaps, the procedure proposed in
  Refs.~\cite{Elgood:2020svt,Elgood:2020mdx,Elgood:2020nls} produces a
  Noether-Wald charge that simply misses terms. However, as shown in
  Ref.~\cite{Mitsios:2021zrn}, the Noether-Wald charge found in this way leads
  to a Smarr formula that also contains magnetic charges and potentials in a
  duality-invariant form, which suggests that nothing is missing from it.} One
can make such a term appear as in Ref.~\cite{Urano:2009xn} but, since the
action of diffeomorphisms on constants is trivial, this does not happen in a
natural way and the physics behind this variation is unclear. The same happens
to coupling constants.

Variations of the cosmological constant are possible in the context of
supergravity/superstring theories, however. In higher-dimensional
supergravity/superstring theories there are higher-rank forms which give
$(d-1)$-form potentials after compactification to $d$ dimensions. These
potentials are dual to constants which are determined dynamically by the
equations of motion of the potentials. The main example is provided by the
3-form potential of 11-dimensional supergravity that gives rise to the
cosmological constant of $\mathcal{N}=8,d=4$ supergravity
\cite{Freund:1980xh,Aurilia:1980xj}, but there are many others.
Ultimately, however, it is expected that all the parameters of lower-dimensional
theories (which can be collected in the embedding tensor) can be explained in
a similar fashion. 

The $(d-1)$-form potentials dual to the constants do have an associated gauge
symmetry. This suggests that the terms proportional to the variations of those
constants in the first law could arise in association with the gauge symmetry
of the dual $(d-1)$-forms. This possibility was first explored in
Refs.~\cite{Teitelboim:1985dp,Brown:1988kg} for the cosmological constant.  We
will explore it in a more systematic way here, using Wald's formalism, for the
cosmological constant and for all the components of the embedding tensor in a
toy model.

In Section~\ref{sec-dualcosmologicalconstant}, we are going to review the
description of the cosmological constant in terms of a $(d-1)$-form potential
in the simplest setting: no matter fields.  We will show that one can recover
the first law of black-hole thermodynamics of
Refs.~\cite{Kastor:2008xb,Kastor:2009wy} using Wald's formalism treating the
gauge symmetry as in
Refs.~\cite{Elgood:2020svt,Elgood:2020mdx,Elgood:2020nls}.\footnote{A
  derivation of the first law in the presence of a cosmological constant
  treated as the conserved charge associated to a $(d-1)$-form potential has
  been carried out in Ref.~\cite{Chernyavsky:2017xwm} using the formalism
  proposed in Ref.~\cite{Hajian:2016iyp}. The Smarr formula was found in
  Ref.~\cite{Hajian:2021hje}.}  Furthermore, we will derive the Smarr formula
using the Komar integral as proposed in
Refs.~\cite{Kastor:2008xb,Kastor:2010gq,Liberati:2015xcp,Jacobson:2018ahi,Ortin:2021ade,Mitsios:2021zrn}.

In Section~\ref{sec-generalexample} we are going to consider a more general
example in 4 dimensions, with two real scalars and a 1-form field coupled to
gravity. The theory is invariant under constant shifts of one of the scalars
and this global symmetry can be gauged using the 1-form and its dual as gauge
fields introducing at the same time two coupling constants that can be
combined into a 2-component embedding tensor. This provides the opportunity to
test the conjectured interpretation of the embedding tensor as a
thermodynamical variable in black-hole physics.\footnote{This conjecture is
  clearly related to and in agreement with the conjecture put forward in
  Ref.~\cite{Hajian:2021hje} that all dimensionful constants in the Lagangian
  contribute to the Smarr formula.}  As it is well known
\cite{Nicolai:2000sc,Nicolai:2001sv,Nicolai:2001sv,deWit:2005ub,deWit:2008ta,deWit:2008gc}
the consistency of this kind of electric/magnetic gaugings demands the
introduction of 2-form fields which, in their turn demand the introduction of
3-form fields etc., giving rise to the so-called \textit{tensor hierarchy}.

Complete tensor hierarchies have been constructed only in a few cases
\cite{Bergshoeff:2009ph,Hartong:2009vc,Fernandez-Melgarejo:2011nso,LassoAndino:2016vmh}
and they show a one-to-one relation between the $(d-2)$-forms and the global
symmetries of the theory that can be gauged (just 1-dimensional in our toy model), between
the $(d-1)$-forms and the deformation constants of the theory (just the 2
components of the embedding tensor in our toy model) and between the $d$-forms
and the constraints satisfied by the deformation constants of the theory (0 in
our toy model). In Section~\ref{sec-generalexample} we will omit the
construction of the tensor hierarchy of the model and we will directly
introduce its fields (the 2 real scalars, the 1-form and its dual, the single
2-form dual to the Noether-Gaillard-Zumino current \cite{Gaillard:1981rj} and
the two 3-forms dual to the coupling constants) and a democratic action based
on the one in Ref.~\cite{deWit:2005ub} in which all of them are present and in
which the components of the embedding tensor are not constants, but functions
which are forced to be constant on-shell.\footnote{This democratic action is a
  true action, as opposite to the democratic action of
  Ref.~\cite{Bergshoeff:2001pv}, which is a \textit{pseudoaction}
  \cite{Bergshoeff:1995sq} whose equations of motion must be supplemented by
  duality constraints to reproduce the equations of motion of the theory.}

Then, in Section~2 we will study the symmetries and define the conserved
charges, including the Noether-Wald charge and the Komar charge, essentially
along the lines of Ref.~\cite{Barnich:2001jy}. We will use the last two to
prove the first law of black hole mechanics, to obtain the Smarr
formula and to study
the role the embedding tensor plays in both of them.

We will discuss our results and their implications in
Section~\ref{sec-discussion}.

Finally, the appendix contains a, not so successful, search for black hole solutions of
our toy model to which our results can be applied. Unfortunately, it is very
difficult to find charged solutions and we only managed to find an embedding
of the Schwarzschild-(a)DS solution into the model. 

\section{Dualizing the cosmological constant}
\label{sec-dualcosmologicalconstant}

The action for pure gravity (described by General Relativity) coupled
to a cosmological constant $\Lambda$ in arbitrary dimension $d$ is

\begin{equation}
\label{eq:action}
S[g_{\mu\nu}]
= 
\frac{1}{16\pi G_{N}^{(d)}}\int 
d^{d}x\sqrt{|g|}\, \left[R(g) -(d-2)\Lambda\right]\,,
\end{equation}

\noindent
and leads to the equations of motion

\begin{equation}
  \label{eq:Einsteinequations}
G_{\mu\nu} +\frac{(d-2)}{2}g_{\mu\nu}\Lambda =0\, ,  
\end{equation}

\noindent
which can be reduced to just

\begin{equation}
R_{\mu\nu} = \Lambda g_{\mu\nu}\,.  
\end{equation}

The dimension-dependent factor of $\Lambda$ in the action has been
chosen so as to arrive to this last equation. On the other hand, in the
conventions that we are using, when $\Lambda$ is positive (negative),
the maximally symmetric solution of the above equations is the (anti-)
De Sitter solution. If we interpret the cosmological term in the
Einstein equations (\ref{eq:Einsteinequations}) as an energy-momentum tensor

\begin{equation}
T_{\mu\nu} = -\frac{(d-2)}{16\pi G_{N}^{(d)}}\Lambda g_{\mu\nu}\,,  
\end{equation}

\noindent
and we compare it with that of a perfect fluid
$-(\rho+p)u_{\mu}u_{\nu} +pg_{\mu\nu}$, we find that the perfect fluid is characterized by

\begin{equation}
\rho = -p = \frac{(d-2)}{16\pi G_{N}^{(d)}}\Lambda\,.  
\end{equation}

In differential-form language, the action (\ref{eq:action}) takes the form

\begin{equation}
\label{eq:diffaction}
  S[e^{a}]
   =
  \frac{(-1)^{d-1}}{16\pi G_{N}^{(d)}} \int
  \left[  \star (e^{a}\wedge e^{b}) \wedge R_{ab}
-(d-2)\star \Lambda
  \right]\,.
\end{equation}

The equations of motion that one obtains from this action, defined by the
variation of the action, up to total derivatives

\begin{equation}
\delta S = \int \mathbf{E}_{a} \wedge \delta e^{a}\,,  
\end{equation}

\noindent
are given by

\begin{equation}
  16\pi G_{N}^{(d)} \mathbf{E}_{a}
  =
  \imath_{a}\star (e^{b}\wedge e^{c}) \wedge R_{bc} -(d-2)\imath_{a}\star \Lambda\,,
\end{equation}

\noindent
and it is not hard to see that they can be rewritten in the form

\begin{equation}
  \label{eq:cosmoEE}
  16\pi G_{N}^{(d)} \mathbf{E}_{a}
  =
(-1)^{d}2\left\{G_{ab}+\frac{(d-2)}{2}g_{ab}\Lambda\right\}\star e^{b}\,,
\end{equation}

\noindent
which provides a check of the equivalence of the actions
Eq.~(\ref{eq:diffaction}) and (\ref{eq:action}).

As it is well known \cite{Aurilia:1980xj},\footnote{See also
  Refs.\cite{Teitelboim:1985dp,Brown:1988kg}.} the cosmological constant
$\Lambda$ can be dualized into a $(d-1)$-form potential that we will denote by
$C$. The dualization can be carried out as follows: first of all, in order to
encompass both the $\Lambda >0$ and $\Lambda <0$ cases, we define

\begin{equation}
\Lambda = \mathrm{sign}\,\Lambda\ \lambda^{2}\,,  
\end{equation}

\noindent
and promote the positive constant $\lambda$ to a function
$\lambda (x)$ that we immediately constrain to be constant by
introducing a Lagrange-multiplier term in the action

\begin{equation}
  \label{eq:democraticlambdaaction}
  \begin{aligned}
    S[e^{a}]  \longrightarrow
    S[e^{a},\lambda,C]
    & = \frac{1}{16\pi G_{N}^{(d)}}
    \int \left[ (-1)^{d-1} \star (e^{a}\wedge e^{b}) \wedge R_{ab}
    \right.
    \\
    & \\
    & \hspace{.5cm}
    \left.
      +(-1)^{d}(d-2)\mathrm{sign}\,\Lambda \star \lambda^{2}
      -C\wedge d\lambda \right]\,,
    \\
    & \\
    & \equiv
    \int \mathbf{L}\,,
  \end{aligned}
\end{equation}

\noindent
with the dual $(d-1)$-form $C$ playing the role of Lagrange multiplier. A
general variation of this action

\begin{equation}
  \label{eq:variationdemocraticlambdaaction}
  \delta S
  =
  \int
  \left\{\mathbf{E}_{a}\wedge \delta e^{a}
    +\mathbf{E}_{\lambda}\delta \lambda
    +\mathbf{E}_{C}\wedge \delta C
   +d\mathbf{\Theta}(\varphi,\delta\varphi)
    \right\}\,,
\end{equation}

\noindent
where $\varphi$ stands for the fields $e^{a},\lambda,C$, gives the equations
of motion and total derivative

\begin{subequations}
  \begin{align}
    \label{eq:Ea}
  16\pi G_{N}^{(d)} \mathbf{E}_{a}
  &  =
    \imath_{a}\star (e^{b}\wedge e^{c}) \wedge R_{bc}
    -(d-2)\, \mathrm{sign}\,\Lambda\,\lambda \imath_{a}\star \lambda\,,
  \\
    & \nonumber \\
    \label{eq:Elambda}
  16\pi G_{N}^{(d)} \mathbf{E}_{\lambda}
  & =
    (-1)^{d-1}
    \left[dC -2(d-2)\mathrm{sign}\,\Lambda\,\star \lambda\right]\,,
  \\
  & \nonumber \\
  16\pi G_{N}^{(d)} \mathbf{E}_{C}
  & =
    (-1)^{d}d\lambda\,,
  \\
  & \nonumber \\
  16\pi G_{N}^{(d)} \mathbf{\Theta}(\varphi,\delta\varphi)
  & =
    -\star (e^{a}\wedge e^{b}) \wedge \delta \omega_{ab}
    +(-1)^{d}C\delta\lambda\,.
\end{align}
\end{subequations}

We can use the equation of motion of $\lambda$

\begin{equation}
  \lambda
  =
  \frac{(-1)^{d-1}\mathrm{sign}\,\Lambda}{2(d-2)} \star dC\,,
\end{equation}

\noindent
to replace $\lambda$ by $G\equiv dC$, the $d$-form field strength of $C$,
arriving at the dual action

\begin{equation}
  \label{eq:dualdiffaction}
  S[e^{a},C]
   =
  \frac{1}{16\pi G_{N}^{(d)}} \int
  \left[ (-1)^{d-1} \star (e^{a}\wedge e^{b}) \wedge R_{ab}
+\frac{\mathrm{sign}\,\Lambda}{4(d-2)} G\star G
\right]\,.
\end{equation}

The variation of the action, up to total derivatives,

\begin{equation}
  \label{eq:variationdualdiffaction}
  \delta S
  =
  \int
  \left\{\mathbf{E}_{a}\wedge \delta e^{a} + \mathbf{E}_{C}\wedge \delta C
    \right\}\,,
\end{equation}

\noindent
gives the following equations of motion

\begin{subequations}
\begin{align}
16\pi G_{N}^{(d)} \mathbf{E}_{a}
&  =
\imath_{a}\star (e^{b}\wedge e^{c}) \wedge R_{bc}
+\frac{(-1)^{d}}{4(d-2)}\mathrm{sign}\,\Lambda\, \imath_{a}G \star G\,,
  \\
  & \nonumber \\
  16\pi G_{N}^{(d)} \mathbf{E}_{C}
  & =
    -\mathrm{sign}\,\Lambda\, d\star G\,.
\end{align}
\end{subequations}

$\mathbf{E}_{C}=0$ is solved by a constant $\star G$. If the constant is
written in the form

\begin{equation}
  \label{eq:onshellstarG}
\star G = (-1)^{d-1} 2(d-2)\mathrm{sign}\, \Lambda\,\,\lambda\,,  
\end{equation}

\noindent
we recover the cosmological Einstein equations Eq.~(\ref{eq:cosmoEE}).

This proves the classical equivalence of the original and the dual
formulations, although the second is slightly more general since the equation
of motion of $C$ can be solved by piecewise constant $\lambda(x)$s whose
discontinuities can be associated to $(d-2)$-brane sources, which couple in a
natural way to the $(d-2)$-form $C$.\footnote{See, for instance,
  Ref.~\cite{Bergshoeff:2001pv}.}

An important difference between $C$ and $\lambda$ is that the former has a
gauge freedom, under which it transforms

\begin{equation}
  \label{eq:gaugetransf}
\delta_{\chi} C = d\chi\,,  
\end{equation}

\noindent
where $\chi$ is an arbitrary $(d-2)$-form. These gauge transformations leave
the field strength $G$ and the dual action Eq.~(\ref{eq:dualdiffaction})
invariant.  The action Eq.~(\ref{eq:democraticlambdaaction}) is also gauge
invariant, but only up to a total derivative. In any case, this invariance is
associated to a conserved charge, apparently not present in the original
system. This charge can be understood in terms of the branes that source $C$
and is directly related to $\lambda$. We study the definition of this
charge in the next section.

Although the actions Eqs.~(\ref{eq:dualdiffaction}) and
(\ref{eq:democraticlambdaaction}) are equivalent, in more complex cases in
which we want to dualize constants that occur in multiple places in the
action, one promotes the constants to fields and adds the Lagrange-multiplier
terms with the dual potentials but one does not take the next step
(eliminating the constants using their equations of motion) because the
resulting actions are too complicated. Thus, one stays with actions similar to
Eq.~(\ref{eq:democraticlambdaaction}) and, therefore, in what follows, we will work
with it.

\subsection{The gauge conserved charge}
\label{sec-conservedcharge}

Under the gauge transformation Eq.~(\ref{eq:gaugetransf}), the action
Eq.~(\ref{eq:democraticlambdaaction}) transforms as

\begin{equation}
  \delta_{\chi} S
  =
  -\frac{1}{16\pi G_{N}^{(d)}} \int d\chi\wedge d\lambda
  =
  \frac{(-1)^{d}}{16\pi G_{N}^{(d)}}
      \int d\left(\lambda d\chi\right)\,.
\end{equation}

\noindent
The total derivative is defined up to the total derivative of a total
derivative, and we have made a choice that we will show is adequate to get a
non-trivial result.

From Eq.~(\ref{eq:variationdemocraticlambdaaction}), instead, upon use of the
Noether identity $d\mathbf{E}_{C}=0$, we get

\begin{equation}
  \delta_{\chi} S
  =
  \int
  -d\left(\mathbf{E}_{C}\wedge \chi \right)
  =
-\frac{(-1)^{d}}{16\pi G_{N}^{(d)}}
  \int
  d\left(d\lambda\wedge \chi\right)\,,
\end{equation}

\noindent
which, together with the previous result leads to the off-shell identity

\begin{equation}
 d\mathbf{J}[\chi]
   =
   0\,,
   \,\,\,\,\,\,
   \text{with}
   \,\,\,\,\,\,
   \mathbf{J}[\chi]
   =
\frac{(-1)^{d-1}}{16\pi G_{N}^{(d)}}
\left(  d\lambda \wedge \chi+ \lambda d\chi\right)\,.
\end{equation}

\noindent
This identity implies that, locally, there must exist a $(d-2)$-form
$\mathbf{Q}[\chi]$ such that

\begin{equation}
  \mathbf{J}[\chi]
  =
  d\mathbf{Q}[\chi]\,,
\end{equation}

\noindent
and it is obvious that 

\begin{equation}
  \mathbf{Q}[\chi]
  =
\frac{(-1)^{d-1}}{16\pi G_{N}^{(d)}} \lambda \chi\,.
\end{equation}

Given a particular solution of the equations of motion $\{e^{a},\lambda,C\}$,
for each inequivalent $(d-2)$-form that preserves it (\textit{i.e.}~for each
harmonic $\chi_{h}$), we can get the conserved charge contained in a closed
$(d-2)$-dimensional surface $\Sigma^{d-2}$ with no boundary by integrating
$\mathbf{Q}[\chi_{h}]$ over it

\begin{equation}
  \label{eq:chargeC}
  \mathcal{Q}[\chi_{h}]
  =
  \frac{(-1)^{d-1}\lambda }{16\pi G_{N}^{(d)}}
  \int_{\Sigma^{d-2}} \chi_{h}\,.
\end{equation}

\noindent
where we have used the fact that on-shell $\lambda $ is constant.

Up to normalization constants this charge is just the volume of $\Sigma^{d-2}$
measured in terms of the volume form $\chi_{h}$. Observe that the value of the
charge does not change under the replacement of $\chi_{h}$ by $\chi_{h}+d e$
for any $(d-3)$-form $e$. Thus, it only depends on the De Rahm cohomological
class of $\chi_{h}$, which is unique (up to normalization) on any compact,
orientable $\Sigma^{d-2}$ with no boundary. It is natural to use the induced
volume form on $\Sigma^{(d-2)}$ that we will denote by
$\Omega_{\Sigma}^{(d-2)}$ and, then,

\begin{equation}
  \label{eq:chargeCnorm}
  \mathcal{Q}
  =
\frac{(-1)^{d-1} \lambda }{16\pi G_{N}^{(d)}}\omega_{\Sigma}\,,
  \,\,\,\,\,
  \text{where}
  \,\,\,\,\,
  \omega_{\Sigma}
  \equiv
  \int_{\Sigma^{d-2}} \Omega_{\Sigma}^{(d-2)}\,.
\end{equation}

Thus, up to numerical constants and the volume $\omega_{\Sigma}$ (not present
in rationalized units) $\lambda$ is the charge carried by $C$.

\subsection{The Noether-Wald charge}
\label{sec-NoetherWaldcharge}

The action Eq.~(\ref{eq:democraticlambdaaction}) is also exactly invariant
under diffeomorphisms and local Lorentz transformations.\footnote{Under
  infinitesimal diffeomorphisms it is invariant only up to a total derivative
  that we will take into account later.} We are interested in the Noether
charge associated to the invariance under diffeomorphisms (Noether-Wald
charge) and, therefore, we start by considering the variation of the action
under diffeomorphisms generated by infinitesimal vector fields $\xi$

\begin{equation}
  \label{eq:deltaxiS}
  \delta_{\xi} S
  =
  \int
  \left\{\mathbf{E}_{a}\wedge \delta_{\xi} e^{a}
    +\mathbf{E}_{C}\wedge \delta_{\xi} C
    +\mathbf{E}_{\lambda}\wedge \delta_{\xi} \lambda
    +d\mathbf{\Theta}(\varphi,\delta_{\xi}\varphi)
    \right\}\,.
\end{equation}

Observe that $\lambda$ must be treated as a scalar field and, therefore

\begin{equation}
  \delta_{\xi}\lambda
  =
  -\imath_{\xi}d\lambda\,.
\end{equation}

\noindent
However, the infinitesimal transformations $\delta_{\xi}$ of $e^{a}$ and $C$
must take into account the gauge freedom of those fields as explained in
Refs.~\cite{Jacobson:2015uqa,Elgood:2020svt,Elgood:2020mdx,Elgood:2020nls} in
such a way that the invariance of the fields under those transformations for a
certain parameter $\xi$ (which we will denote by $k$) is a gauge-invariant
statement. Since, in particular, $\delta_{k}$ must leave invariant the metric,
$k$ is always a Killing vector.  The transformations $\delta_{\xi}e^{a}$ and
$\delta_{\xi}C$ are combinations of standard Lie derivatives and
$\xi$-dependent ``compensating'' gauge transformations

\begin{subequations}
  \begin{align}
    \sigma_{\xi}^{ab}
    & =
  \imath_{\xi}\omega^{ab} -P_{\xi}{}^{ab}\, ,
  \,,
    \\
    & \nonumber \\
    \label{eq:compensatingchi}
    \chi_{\xi}
    & =
      \imath_{\xi}C -P_{\xi}\,,
  \end{align}
\end{subequations}

\noindent
where

\begin{equation}
  \label{eq:Pxiab}
  P_{\xi}{}^{ab} \equiv \nabla^{[a}\xi^{b]}\,,
\end{equation}

\noindent
is the (scalar) Lorentz momentum map, which satisfies  for $\xi=k$

\begin{equation}
  \mathcal{D}P_{k}{}^{ab}
  =
  -\imath_{k}R^{ab}\,.
\end{equation}

\noindent
On the other hand, $P_{\xi}$ is the $(d-2)$-form momentum map associated to
$C$, which is such that, for $\xi=k$

\begin{equation}
  \label{eq:Gmomentummap}
  dP_{k}= -\imath_{k}G\,.
\end{equation}

After some massaging, we can write the transformations in the form 

\begin{subequations}
  \label{eq:deltaxi}
  \begin{align}
    \delta_{\xi}e^{a}
    & =
      -(\mathcal{D}\xi^{a}+P_{\xi}{}^{a}{}_{b}e^{b})\,,
    \\
    & \nonumber \\
    \delta_{\xi}\omega^{ab}
    & =
      -(\imath_{\xi}R^{ab}+\mathcal{D}P_{\xi}{}^{ab})\,,
    \\
    & \nonumber \\
    \delta_{\xi}C
    & =
    -(\imath_{\xi}G +dP_{\xi})\,.
  \end{align}
\end{subequations}

The definitions of the momentum maps ensure that
$\delta_{k}e^{a}=\delta_{k}\omega^{ab}=\delta_{k}C=0$ in a gauge-invariant
fashion.

Observe that, on-shell,
\begin{equation}
  \imath_{k}G
  =
  (-1)^{d-1}2(d-2)\,\mathrm{sign}\,\Lambda\,\,\lambda \star\hat{k}\,,  
\end{equation}

\noindent
where $\hat{k}=k_{\mu}dx^{\mu}$ if $k=k^{\mu}\partial_{\mu}$.\footnote{With
  our conventions,
  \begin{equation}
  \imath_{k}\star \mathbb{I} = (-1)^{d-1}\star \hat{k}\,.  
  \end{equation}
}Then,

\begin{equation}
  \label{eq:Gmomentummap2}
  P_{k}
  =
  (-1)^{d}2(d-2)\mathrm{sign}\,\Lambda\,\,\lambda\, \omega_{k}\,,
\end{equation}

\noindent
where the $(d-2)$-form $\omega_{k}$ is the $d$-dimensional generalization of
the \textit{Killing co-potential} introduced in Ref.~\cite{Kastor:2008xb}
defined by

\begin{equation}
d\omega_{k}=\star \hat{k}\,.  
\end{equation}

Although the existence of $P_{k}$ and, hence, of $\omega_{k}$ was
initially guaranteed by the invariance of $G$ under $\delta_{k}$, we
see here that it is also related to $k$ being a Killing vector. Since
on-shell $G$ is, up to constants, the metric volume form, these two
facts are obviously related.

Substituting the transformations Eqs.~(\ref{eq:deltaxi}) into
Eq.~(\ref{eq:deltaxiS}), using the Noether identities associated to
the symmetries and performing simple manipulations we arrive at

\begin{equation}
  \delta_{\xi}S
  =
    \int d \left\{ \mathbf{\Theta}(\varphi,\delta_{\xi}\varphi)
+(-1)^{d} \mathbf{E}_{a}\xi^{a} +\mathbf{E}_{C}\wedge P_{\xi}
\right\}
\equiv
\int d\mathbf{\Theta}'\,.
\end{equation}

Now we must take into account that the action
Eq.~(\ref{eq:democraticlambdaaction}) is invariant under diffeomorphisms and
gauge transformations up to total derivatives:

\begin{equation}
  \delta_{\xi}S
  =
  \int d \left\{ -\imath_{\xi}\mathbf{L}
    +\frac{(-1)^{d-1}}{16\pi G_{N}^{(d)}}
    \left[
      d\lambda \wedge \imath_{\xi}C
      +
      \lambda dP_{\xi}\right]
    \right\}\, ,
\end{equation}

\noindent
where we have used the explicit form of the compensating $\delta_{\chi}\xi$
transformation Eq.~(\ref{eq:compensatingchi}) and the freedom that we have to
add total derivatives of total derivatives to obtain a convenient expression.

We arrive at the off-shell identity

\begin{equation}
d\mathbf{J}[\xi]=0\,,  
\end{equation}

\noindent
where

\begin{equation}
  \label{eq:Jxidef}
  \mathbf{J}[\xi]
  \equiv
\mathbf{\Theta}'  
+\imath_{\xi}\mathbf{L}
 +\frac{(-1)^{d}}{16\pi G_{N}^{(d)}}
 \left[d\lambda \wedge \imath_{\xi}C+\lambda dP_{\xi}\right]\,,
\end{equation}

\noindent
and it is not difficult to see that

\begin{equation}
  \label{eq:JdQ}
  \mathbf{J}[\xi]
  =
  d \mathbf{Q}[\xi]\,,
\end{equation}

\noindent
where the Wald-Noether $(d-2)$-form $\mathbf{Q}[\xi]$ is given by 

\begin{equation}
  \label{eq:Qxi}
  \mathbf{Q}[\xi]
  \equiv
  \frac{(-1)^{d}}{16\pi G_{N}^{(d)}}
  \left\{\star (e^{a}\wedge e^{b})P_{\xi\, ab}
      +\lambda P_{\xi}\right\}\,.  
\end{equation}

\subsection{The generalized, restricted, zeroth law}
\label{sec-zerothlaw}

A crucial ingredient in the proof of the first law of black-hole mechanics
along the lines of
Refs.~Refs.~\cite{Jacobson:2015uqa,Elgood:2020svt,Elgood:2020mdx,Elgood:2020nls}
are the \textit{generalized, restricted zeroth laws}. These laws are called
``generalized'' because they generalize the standard zeroth law of black-hole
mechanics stating that the surface temperature $\kappa$ is constant over the
event horizon $\mathcal{H}$ to other thermodynamical potentials such as the
electrostatic black-hole potential. On the other hand, they are are called
``restricted'' because their validity is restricted to the bifurcation surface
$\mathcal{BH}$\footnote{One could use the arguments of
  Ref.~\cite{Jacobson:1993vj} to extend their validity to the complete event
  horizon, though.} and because, rather than stating the constancy of a scalar
quantity, they just state the closedness of a given differential form over
$\mathcal{BH}$.\footnote{Actually, when dealing with forms of rank higher than
  1 (1 would correspond to an electromagnetic field), it is not clear which
  other covariant statement could play the role of zeroth law.} This is enough
for our purposes, though.

Thus, at this point we are going to focus on solutions of the action
Eq.~(\ref{eq:democraticlambdaaction}) which describe stationary black-hole
spacetimes with a cosmological constant determined by the value of $\lambda$,
with bifurcate event horizons that coincide with the Killing horizon
associated to a certain asymptotically timelike Killing vector $k$. By
definition, $k$ vanishes over the bifurcation surface which we denote by
$\mathcal{BH}$

\begin{equation}
k \stackrel{\mathcal{BH}}{=} 0\,.  
\end{equation}

\noindent
Thus, if all fields are regular over the horizon, it is clear that the inner
products of their field strengths with $k$ must vanish on $\mathcal{BH}$ :

\begin{subequations}
  \begin{align}
    \label{eq:ikG=0}
    \imath_{k}G & \stackrel{\mathcal{BH}}{=} 0\,,
    \\
    & \nonumber \\
    \label{eq:ikR=0}
    \imath_{k}R^{a}{}_{b} & \stackrel{\mathcal{BH}}{=} 0\,. 
  \end{align}
\end{subequations}

Let us consider the first of these properties. According to the definition
Eq.~(\ref{eq:Gmomentummap}), the $(d-2)$-form $P_{k}$ is closed on
$\mathcal{BH}$. Being a $(d-2)$-form on $\mathcal{BH}$, it must be
proportional to the induced volume form of $\mathcal{BH}$, $\Omega_{\cal BH}$:

\begin{equation}
P_{k} = f \Omega_{\cal BH}\,.  
\end{equation}

\noindent
Then, the closedness of $P_{k}$ implies that the coefficient $f$ is a
constant.  This statement is one of the generalized, restricted zeroth laws of
black-hole mechanics that have been used in
Refs.~\cite{Copsey:2005se,Compere:2007vx,Jacobson:2015uqa,Prabhu:2015vua,Elgood:2020svt,Elgood:2020mdx,Elgood:2020nls}
to prove the first law. If we normalize the volume form so that its integral
is equal to $1$, $f$ will be proportional to the volume of $\mathcal{BH}$.  This is
the thermodynamical potential (``volume'') associated to the thermodynamical
variable $\lambda$ (``pressure'').  In order to make contact with the
conventions of Ref.~\cite{Ortin:2021ade}, it is more convenient to use the
Killing co-potential $(d-2)$-form $\omega_{k}$, which due to
Eq.~(\ref{eq:Gmomentummap2}), the must also be closed on $\mathcal{BH}$
on-shell and, therefore, proportional to the volume form. Thus, we define the
volume $\Theta_{\lambda}$ by\footnote{Apart from the sign, there is another
  difference with most of the literature in black-hole chemistry: this volume
  is proportional to $\lambda$, which is natural for a potential, but,
  perhaps, not for a volume.}

\begin{equation}
  \label{eq:Thetalambdadef}
\frac{P_{k}}{16\pi G_{N}^{(d)}} = (-1)^{d-1}
  \Theta_{\lambda} \omega_{k}/V_{k}\,,
  \,\,\,\,\,\,
  \text{with}
  \,\,\,\,\,\,
  V_{k} \equiv  \int_{\cal BH}\omega_{k}\,,
  \,\,\,\,\,
  \Theta_{\lambda} = -\mathrm{sign}\,\Lambda\ \frac{(d-2)\lambda\, V_{k}}{8\pi G_{N}^{(d)}}\,,
\end{equation}

\noindent
so that the volume $\Theta_{\lambda}$ is positive for aDS black holes (sign$\Lambda <0$).

Following Ref.~\cite{Ortin:2021ade}, $\Theta_{\lambda} $ can written as

\begin{equation}
  \label{eq:Thetalambda}
  \Theta_{\lambda}
  = \frac{(-1)^{d}}{16\pi G_{N}^{(d)}}
  \int_{\cal B}\imath_{k}\star \frac{\partial \mathcal{V}}{\partial
    \lambda}\,,
  \,\,\,\,\,\,
  \text{with}
  \,\,\,\,\,\,
  \mathcal{V} \equiv (d-2) \mathrm{sign}\,\Lambda\ \lambda^{2}\,,
\end{equation}

\noindent
where $\mathcal{B}$ is a ball whose radius is that of the horizon and whose
boundary is $\mathcal{BH}$. This is an expression that we will generalize
later on.

The property Eq.~(\ref{eq:ikR=0}) is related to the standard zeroth law of
black-hole mechanics because it implies

\begin{equation}
  \label{eq:DPkab=0}
\mathcal{D}P_{k\, ab}  \stackrel{\mathcal{BH}}{=} 0\,,  
\end{equation}

\noindent
and because, on the bifurcation surface

\begin{equation}
  \label{eq:Pkappan}
P_{k\, ab}  \stackrel{\mathcal{BH}}{=} \kappa n_{ab}\,,  
\end{equation}

\noindent
where $n_{ab}$ is the binormal to $\mathcal{BH}$, with the normalization
$n^{ab}n_{ab}=-2$. Since $\kappa$ is constant according to the zeroth law,
$n_{ab}$ must be covariantly constant on $\mathcal{BH}$. We do not have an
independent proof of this property, which is of purely geometric nature. With
this proof in hand, the zeroth law on $\mathcal{BH}$
($d\kappa \stackrel{\cal BH}{=}0$) would be a consequence of
Eq.~(\ref{eq:DPkab=0}). All zeroth laws (generalized or not) would follow the
same pattern since they would state that the coefficients of the expansion of
certain closed (or covariantly-closed) forms in a properly defined and
normalized basis are constant as in
Refs.~\cite{Copsey:2005se,Compere:2007vx,Jacobson:2015uqa,Prabhu:2015vua,Elgood:2020svt,Elgood:2020mdx,Elgood:2020nls}.

\subsection{Komar integral and Smarr formula}
\label{sec-Komar}

Before we use the Noether-Wald charge and the restricted, generalized second
laws to prove the first law of black-hole mechanics \cite{Bardeen:1973gs}, it
is useful to test our results constructing a Komar integral
\cite{Komar:1958wp} following
Refs.~\cite{Kastor:2008xb,Kastor:2010gq,Liberati:2015xcp,Jacobson:2018ahi,Ortin:2021ade,Mitsios:2021zrn}
and using it to derive a Smarr formula \cite{Smarr:1972kt} that can be tested
in actual black-hole solutions.

On-shell\footnote{Here we use the symbol $\doteq$ for identities that only
  hold on-shell.} and for a Killing vector $k$ that generates a symmetry of
the whole field configuration, the Noether-Wald current defined in
Eq.~(\ref{eq:Jxidef}) satisfies

\begin{equation}
  \mathbf{J}[k]
  \,\,\doteq\,\,
  \imath_{k}\mathbf{L}
    +\frac{(-1)^{d}}{16\pi G_{N}^{(d)}}\lambda dP_{k}\,.  
\end{equation}

On the other hand, $\mathbf{J}[k]$ satisfies Eq.~(\ref{eq:JdQ}) off-shell with
$\xi=k$, which implies that 

\begin{equation}
  d\mathbf{Q}[k]
  -\imath_{k}\mathbf{L}
  -\frac{(-1)^{d}}{16\pi G_{N}^{(d)}}\lambda
dP_{k}
=
0\,.  
\end{equation}

We can write a Komar integral with volume terms, as in
Ref.~\cite{Liberati:2015xcp}, or we can take a step further and rewrite the
last two terms as total derivatives
Refs.~\cite{Ortin:2021ade,Mitsios:2021zrn}. This is trivial for the second
term. As for the first additional term, if $k$ generates a symmetry of the
whole field configuration,

\begin{equation}
0\,\,\doteq\,\,\pounds_{k}\mathbf{L}= d\imath_{k}\mathbf{L}\,,  
\end{equation}

\noindent
and $\imath_{k}\mathbf{L}$ must be locally exact. Therefore, there must exist a
$(d-2)$-form $\varpi_{k}$ such that 

\begin{equation}
  d \varpi_{k} \,\,\doteq\,\, \imath_{k}\mathbf{L}
  +\frac{(-1)^{d}}{16\pi G_{N}^{(d)}}\lambda dP_{k}\,, 
\end{equation}

\noindent
which leads to the identity

\begin{equation}
  \label{eq:Komaridentity}
d\left\{\mathbf{Q}[k]-\varpi_{k}\right\}\doteq0\,.  
\end{equation}

\noindent
Then, the Komar integral over the codimension-2 surface $\Sigma^{d-2}$ can be
defined as the integral over the \textit{Komar charge}
$-\left\{\mathbf{Q}[k]-\varpi_{k}\right\}$ \cite{Ortin:2021ade}

\begin{equation}
  \label{eq:Komarintegral}
  \mathcal{K}(\Sigma^{d-2})
  =
  -\int_{\Sigma^{d-2}}\left\{\mathbf{Q}[k]-\varpi_{k}\right\}\,.
\end{equation}

In order to determine $\varpi_{k}$, we first calculate the on-shell value of the
Lagrangian density: tracing over the Einstein equations (\ref{eq:Ea})

\begin{equation}
  \begin{aligned}
    e^{a}\wedge\mathbf{E}_{a}
    & =
    (d-2)\star (e^{b}\wedge e^{c}) \wedge R_{bc}
    -(d-2)d\, \mathrm{sign}\,\Lambda\,\lambda \star \lambda
    \\
    & \\
    & =
    (-1)^{d-1}(d-2)\left\{\mathbf{L}
      +\frac{(-1)^{d}\,\mathrm{sign}\,\Lambda}{8\pi G_{N}^{(d)}}
       \star\lambda^{2}\right\}\,,
  \end{aligned}
\end{equation}

\noindent
so 

\begin{equation}
  \mathbf{L}\,\,
  \doteq\,\,
  \frac{(-1)^{d-1}\,\mathrm{sign}\,\Lambda}{8\pi G_{N}^{(d)}}
       \star\lambda^{2}\,.
\end{equation}

Now, using the equation of motion of $\lambda$, Eq.~(\ref{eq:Elambda}), to
replace $\star \lambda$ by $G$ and the definition of the momentum map $P_{k}$
in Eq.~(\ref{eq:Gmomentummap}) to replace $\imath_{k}G$ by $-dP_{k}$, we get

\begin{equation}
  \imath_{k}\mathbf{L}
  \,\,\doteq\,\,
  d\left\{\frac{(-1)^{d}\, \lambda P_{k}}{(d-2)16\pi G_{N}^{(d)}}\right\}\,,
\end{equation}

\noindent
and

\begin{equation}
  \varpi_{k}
  =
  \frac{(-1)^{d}(d-1)\, \lambda P_{k}}{(d-2)16\pi G_{N}^{(d)}}\,.
\end{equation}

The Komar integral is, then

\begin{equation}
  \mathcal{K}(\Sigma^{d-2})
  =
  \frac{(-1)^{d-1}}{16\pi G_{N}^{(d)}}
  \int_{\Sigma^{d-2}}
  \left\{\star(e^{a}\wedge e^{b})P_{k\, ab} -\frac{\lambda P_{k}}{(d-2)}\right\}\,.
\end{equation}

Let us consider the anti-De Sitter case ($\mathrm{sgn}\, \Lambda <0
$): if integrate the exterior derivative of the integrand over a hypersurface
whose boundary is the union of a spatial section of a stationary black-hole
Killing horizon (the bifurcation surface, $\mathcal{BH}$, for the sake of
convenience) and spatial infinity, $S^{d-2}_{\infty}$, Stokes' theorem tells
us that

\begin{equation}
  \label{eq:Komaridentity2}
  \mathcal{K}(\mathcal{BH})
  =
 \mathcal{K}(S^{d-2}_{\infty})\,.
\end{equation}

For the sake of simplicity, let us consider a static, spherically symmetric
black-hole solution: the Schwarzschild-aDS-Tangherlini solution
\cite{Tangherlini:1963bw}, whose metric is given by

\begin{equation}
  ds^{2} = Wdt^{2}-W^{-1}dr^{2} -r^{2}d\Omega^{2}_{(d-2)}\,,
  \,\,\,\,\,\,
  \text{with}
  \,\,\,\,\,\,
  W= 1-\frac{2m}{r^{d-3}}+\frac{|\Lambda|}{d-1}r^{2}\,, 
\end{equation}

\noindent
where

\begin{equation}
m = \frac{8\pi G_{N}^{(d)}M}{(d-2)\omega_{(d-2)}}\,,  
\end{equation}

\noindent
$\omega_{(d-2)}$ being the volume of the unit, round, $(d-2)$-sphere, and $M$
the ADM mass.

The event horizon of this solution is placed at some value $r_{h}$ at which
$W(r_{h})=0$. The Hawking temperature and Bekenstein-Hawking entropy can be
expressed in terms of $r_{h}$ even if its value cannot be determined
explicitly. They are given, respectively, by \cite{Kastor:2009wy}

\begin{subequations}
  \begin{align}
    T
    & =
      \frac{1}{2\pi (d-1)r_{h}^{d-2}}
      \left[(d-1)(d-3)m+|\Lambda| r_{h}^{d-1}\right]\,,
    \\
    & \nonumber \\
    S
    & =
      \frac{\omega_{(d-2)}r_{h}^{d-2}}{4G_{N}^{(d)}}\,,
  \end{align}
\end{subequations}

\noindent
and the product $ST$ leads to the Smarr formula

\begin{equation}
  \label{eq:SmarrSaDST}
\frac{(d-3)}{(d-2)}M
=
ST
 -\frac{1}{8\pi G_{N}^{(d)}} \frac{\omega_{(d-2)}r_{h}^{d-1}}{(d-1)}
      |\Lambda|\,.
\end{equation}

We can evaluate the Komar integral on a constant $r$ surface $S^{d-2}_{r}$

\begin{equation}
  \mathcal{K}(S^{d-2}_{r})
  =
  \frac{\omega_{(d-2)}}{16\pi G_{N}^{(d)}} \left\{r^{d-2}W'
    -2r^{d-1}\frac{|\Lambda|}{(d-1)}  \right\}\,.
\end{equation}

\noindent
At infinity, the second term cancels a divergent term coming from $W'$ and we
get the left-hand side of the Smarr relation Eq.~(\ref{eq:SmarrSaDST}).  At
the horizon, the first term gives directly $ST$ and we get the right-hand side
of Eq.~(\ref{eq:SmarrSaDST}) and we recover the complete Smarr relation from
the Komar integral.

Observe that, since the restricted, generalized, zeroth law guarantees that,
over the bifurcation surface, $P_{k}$ is a constant times the volume form, in
general we can take that constant outside of the Komar integral
$\mathcal{K}(\mathcal{BH})$. In this simple case, also $\lambda$ can be taken
out of the integral, but in more general cases, only the constant defining the
momentum map can be taken outside the integral. The integral of $\lambda$ is
also the integral of $\star G$ on-shell, which gives, up to normalization
constants, the associated charge. 

Finally, using the definition of $\Theta_{\lambda}$ in
Eq.~(\ref{eq:Thetalambda}) and

\begin{equation}
V_{k}= \frac{r_{h}^{d-1}\omega_{(d-2)}}{d-1}\,,  
\end{equation}

\noindent
the Smarr formula can be written in the form

\begin{equation}
  \label{eq:SmarrSaDST-2}
  (d-3)M
=
(d-2)ST-\Theta_{\lambda}\lambda \,,
\end{equation}

\noindent
which is the form that follows from the usual scaling and homogeneity
arguments.\footnote{$1/\lambda$ has dimensions of length.}

\subsection{The first law and black-hole chemistry}
\label{sec-firstlaw}

We are ready to proof the first law of black-hole mechanics in this theory
using Wald's formalism \cite{Lee:1990nz,Wald:1993nt,Wald:1993nt}.

We consider field configurations that describe asymptotically flat,
stationary, black-hole spacetimes admitting a timelike Killing vector $k$
whose bifurcate Killing horizon coincides with the black hole's event horizon
$\mathcal{H}$. $k$, then, will be given by a linear combination with constant
coefficients $\Omega^{n}$ of the timelike Killing vector associated to
stationarity, $t^{\mu}\partial_{\mu}$ and the $[\tfrac{1}{2}(d-1)]$ generators
of inequivalent rotations in $d$ spacetime dimensions
$\phi_{n}^{\mu}\partial_{\mu}$

\begin{equation}
  k^{\mu} = t^{\mu} +\Omega^{n}\phi_{n}^{\mu}\, .  
\end{equation}

\noindent
The constant coefficients $\Omega^{n}$ are the angular velocities of
the horizon.

The starting point of the proof is the fundamental relation
\cite{Lee:1990nz,Wald:1993nt,Iyer:1994ys}

\begin{equation}
  d\left( \delta \mathbf{Q}[k]+\imath_{k}\mathbf{\Theta}' \right)
  =
  0\,,
\end{equation}

\noindent
valid for on-shell field configurations $\varphi$ satisfying the equations of
motion and perturbations of the fields $\delta\varphi$ satisfying the
linearized equations of motion.

We are going to integrate this relation over the hypersurface $\Sigma$ defined
as the space bounded by infinity and the bifurcation sphere $\mathcal{BH}$ on
which $k=0$. Therefore, its boundary, $\delta\Sigma$, has two disconnected
pieces: a $(d-2)$-sphere at infinity, S$^{d-2}_{\infty}$, and the bifurcation
sphere $\mathcal{BH}$.  Using Stokes theorem and taking into account that
$k=0$ on $\mathcal{BH}$, we obtain the relation

\begin{equation}
  -\delta \int_{\mathcal{BH}}   \mathbf{Q}[k]
  =
  -\int_{\mathrm{S}^{d-2}_{\infty}}
  \left( \delta \mathbf{Q}[k]+\imath_{k}\mathbf{\Theta}' \right)\,,
\end{equation}

\noindent
where we have added conventional minus signs that take into account the minus
sign in our definitions of the variations of the fields under diffeomorphisms.

As explained in Ref.~\cite{Iyer:1994ys,Compere:2007vx}, the right-hand side
can be identified with $\delta M -\Omega^{n}\delta J_{n}$, where $M$ is the
total mass of the black-hole spacetime and $J_{n}$ are the independent
components of the angular momentum.

Using the explicit form of the Noether-Wald charge Eq.~(\ref{eq:Qxi}) 

\begin{equation}
  \label{eq:deltaintQ1}
 -\delta \int_{\mathcal{BH}} \mathbf{Q}[k]
 =
 \frac{(-1)^{d-1}}{16\pi G_{N}^{(d)}}
\delta \int_{\mathcal{BH}}
     \star (e^{a}\wedge e^{b}) P_{k\, ab}
     +\frac{(-1)^{d-1}}{16\pi G_{N}^{(d)}}  \delta \lambda
     \int_{\mathcal{BH}} P_{k}\,.
\end{equation}

The right-hand side of this identity is expected to be of the form
$T\delta S+\Phi\delta\mathcal{Q}$ for some charges $\mathcal{Q}$ and
potentials $\Phi$ and/or ``pressures'' $\vartheta$ and ``volumes''
$\Theta_{\vartheta}$.  In this expression $\lambda$ plays the role of charge
or pressure, while $\Theta_{\lambda}$, defined in
Eq.~(\ref{eq:Thetalambdadef}), plays the role of conjugate potential or
volume.  Using Eqs.~(\ref{eq:onshellstarG}) and (\ref{eq:Thetalambdadef})
($\mathrm{sign}\,\Lambda=-1$)

\begin{equation}
\frac{(-1)^{d-1}}{16\pi G_{N}^{(d)}}  \delta \lambda  \int_{\mathcal{BH}} P_{k}
=
\Theta_{\lambda} \delta \lambda\,.
\end{equation}

Using also Eq.~(\ref{eq:Pkappan}) we arrive at

\begin{equation}
  \delta M
  =
  T\delta S
  +\Omega^{n}\delta J_{n}
 +\Theta_{\lambda} \delta \lambda\,.
\end{equation}

The (unconventional, in black-hole chemistry literature) factor of $\lambda$
present in the definition of $\Theta_{\lambda}$ can be absorbed in
$\delta$. However, when the cosmological constant arises as the square of
another, more fundamental constant, as in gauged supergravity, this form of
the first law is more natural. Also, in these theories, the coupling constant
is often associated to $(d-1)$-form potentials coming from higher dimensions
\cite{Freund:1980xh}.
\par
A a matter of fact, it is always possible to introduce a
$(d-1)$-form potential dual to the coupling constants, masses or any other
parameters occurring in the action. These $(d-1)$-forms are part of what is
known as the \textit{tensor hierarchy} of the theory. At the level of the
action they can always be introduced in the same way we introduced the
potential dual to the cosmological constant: promoting first the parameters to
fields and introducing the dual $(d-1)$-form potentials as Lagrange
multipliers that constrain the fields/parameters to be constant. Intuitively,
we expect terms in the first law and Smarr formula associated to all those
$(d-1)$ potentials and, henceforth, to all those coupling constants, masses
and other parameters.

In the next section we are going to consider a very simple
model inspired by gauged supergravity, in which we can test these ideas.

\section{A more general example}
\label{sec-generalexample}

In this section we want to consider a more general model which essentially
describes two scalars $\phi^{1},\phi^{2}$ and a 1-form field $A$ coupled to
gravity, represented by the Vierbein $e^{a}$, in $d=4$. In this model, the
invariance under constant shifts of $\phi^{2}$ has been gauged using a
combination of the 1-form $A$ and its dual $\tilde{A}$ as gauge fields with
two coupling constants $\vartheta$ and $\vartheta$. By consistency, it is
necessary to introduce a 2-form $B$ which can be taken to be the dual of the
Noether current $j$ associated to the invariance under constant shifts of
$\phi^{2}$, so no additional degrees of freedom are added to the
theory. Actually, one can write an action for all these fields which gives the
expected equations of motion plus the duality relations between $j$ and $B$
and between $A$ and $\tilde{A}$ (see Ref.~\cite{deWit:2005ub}).

One can go further, dualizing the two coupling constants into 2 3-forms $C$
and $\tilde{C}$ as we have done with the cosmological constant in the previous
section, completing the \textit{tensor hierarchy} as in
Refs.~\cite{Bergshoeff:2009ph,Hartong:2009vc}. Again, this introduces no new
local degrees of freedom.

Before introducing the action that describes this system, we introduce some
notation: the coupling constants, their dual 3-forms, the 1-form and its dual
and the the 0- and 2-form gauge parameters are collected in symplectic vectors
$\vartheta_{M},C^{M},A^{M},\sigma^{M},\chi^{M}$ as follows:

\begin{equation}
  \begin{array}{rclrclrcl}
    \left(\vartheta_{M}\right)
  & \equiv & 
  \left(\vartheta, \tilde{\vartheta} \right)\,,
             \hspace{.6cm}
             &
\left(A^{M}\right)
  & \equiv & 
  \left(
    \begin{array}{c}
      A \\ \tilde{A} \\
    \end{array}
  \right)\,,
    \hspace{.6cm}
    &
  \left(C^{M}\right)
 & \equiv &
  \left(
    \begin{array}{c}
      C \\ \tilde{C} \\
    \end{array}
    \right)\,,
    \\
    & & & & & & & & \\
  \left( \vartheta^{M}\right)
 & = &
  \left(
    \begin{array}{r}
      -\tilde{\vartheta} \\ \vartheta\\
    \end{array}
    \right)\,,
    \hspace{.6cm}
    &
  \left(\sigma^{M}\right)
  & \equiv &
  \left(
    \begin{array}{c}
      \sigma \\ \tilde{\sigma} \\
    \end{array}
  \right)\,,
    \hspace{.6cm}
    &
  \left(\chi^{M}\right)
  & \equiv &
  \left(
    \begin{array}{c}
      \chi \\ \tilde{\chi} \\
    \end{array}
    \right)\,.
  \end{array}
\end{equation}

The field strengths are defined as

\begin{subequations}
  \begin{align}
  \mathcal{D}\phi^{2}
  & \equiv
    d\phi^{2} -\vartheta_{M}A^{M}\,,
    \\
    & \nonumber \\
    F^{M}
    & =
     dA^{M}+\vartheta^{M}B\,,      
    \\
    & \nonumber \\
    H
    & =
     dB\,,
    \\
    & \nonumber \\
    G^{M}
    & \equiv
      dC^{M} +A^{M}\wedge \star j +\tfrac{1}{2}\vartheta^{M} B\wedge B
      +\delta^{M, 2}B\wedge (\star F-\tilde{F})\,,
  \end{align}
\end{subequations}

\noindent
where

\begin{equation}
  j
  \equiv
  (\phi^{1})^{2}\mathcal{D}\phi^{2}\,,  
\end{equation}

\noindent
and $\delta^{M\,\tilde{.}}$ is 1 for $\tilde{C}$ and zero for $C$.  Under the
gauge transformations

\begin{subequations}
  \begin{align}
    \delta \phi^{2}
    & =
     \vartheta_{M}\sigma^{M} \,,
    \\
    & \nonumber \\
    \label{eq:deltaAM}
    \delta A^{M}
    & =
      d\sigma^{M} -\vartheta^{M}\Lambda\,,
    \\
    & \nonumber \\
    \label{eq:deltaB}
    \delta B
    & =
      d\Lambda\,,
    \\
    & \nonumber \\
    \label{eq:deltaC}
    \delta C^{M}
    & =
      d\chi^{M} -\sigma^{M}\star j -\vartheta^{M}\Lambda\wedge B
      -\delta^{M, 2}\Lambda\wedge (\star F -\tilde{F})\,,
  \end{align}
\end{subequations}

\noindent
the above field strengths are on-shell invariant only. More precisely,
$\mathcal{D}\phi^{2},F^{M}$ and $H$ are gauge invariant up to terms
proportional to $d\vartheta_{M}$ while the 4-form fields strengths $G^{M}$ are
gauge invariant up to terms proportional to $d\vartheta_{M}$ and to equations
of motion that establish relations of duality among the fields. Nevertheless,
the action that we are going to use is off-shell gauge invariant, up to total
derivatives.

Suppressing the normalization factor $(16\pi G_{N}^{(4)})^{-1}$, for the
moment, the action takes the form

\begin{equation}
  \begin{aligned}
    \label{eq:actionaxample}
    S
    & =
\int
\left\{ -\star (e^{a}\wedge e^{b}) \wedge R_{ab}
+\tfrac{1}{2}d\phi^{1}\wedge \star d\phi^{1}
+\tfrac{1}{2} (\phi^{1})^{2}\mathcal{D}\phi^{2}\wedge \star \mathcal{D}\phi^{2}
\right.
\\
& \\
& \hspace{.5cm}
\left.
+\tfrac{1}{2} F\wedge \star F
+\tilde{\vartheta}B\wedge \left(\tilde{F}-\tfrac{1}{2}\vartheta B\right)
-C^{M}\wedge d\vartheta_{M}
+\star V(\phi)
\right\}\,,
  \end{aligned}
\end{equation}

\noindent
where the potential is assumed to be a function of $\phi^{1}$ only and of the
two coupling constants $\vartheta,\tilde{\vartheta}$ which are needed for
dimensional reasons; for the same reason they must appear in it
quadratically, so that the potential is a homogeneous function of
degree two, {\em i.e.\/} the potential satisfies

\begin{equation}
 \label{eq:PotHomFnctDeg2}
 \vartheta_{M}\frac{\partial V}{\partial\vartheta_{M}}
 \ =\ 2\ V \,.  
\end{equation}

The second and third terms in the second line of the action are gong to be referred
to as ``additional''; they are topological and do not contain kinetic terms.

The equations of motion are defined by the general variation of the action

\begin{equation}
  \label{eq:generalvariation}
  \begin{aligned}
  \delta S
 & =
  \int \left\{
    \mathbf{E}_{a}\wedge \delta e^{a}
    +\mathbf{E}_{1}\delta\phi^{1}
    +\mathbf{E}_{2} \delta\phi^{2}
    +\mathbf{E}_{A^{M}}\wedge \delta A^{M}
    +\mathbf{E}_{B}\wedge \delta B
    +\mathbf{E}_{C^{M}}\wedge \delta C^{M}
  \right.
  \\
  & \\
  & \hspace{.5cm}
  \left.
    +\mathbf{E}_{\vartheta_{M}}\wedge \delta \vartheta_{M}
     +d\mathbf{\Theta}(\varphi,\delta\varphi)
  \right\}\,.
\end{aligned}
\end{equation}

The equations of the 3-forms are just

\begin{equation}
    \mathbf{E}_{C^{M}}
    =
      d\vartheta_{M}\,,
\end{equation}

\noindent
so that the $\vartheta_{M}$ are (piecewise) constant on-shell, as intended.

The Einstein equations only involve the field strengths of the fundamental
fields $\phi^{1},\phi^{2}$ and $A$ because the additional terms are all
topological:

\begin{equation}
  \label{eq:Eaexample}
  \begin{aligned}
  \mathbf{E}_{a}
  & =
    \imath_{a}\star (e^{c}\wedge e^{d})\wedge R_{cd}
      +\tfrac{1}{2}
    \left(\imath_{a}d\phi^{1}\star d\phi^{1}
    +d\phi^{1}\wedge \imath_{a}\star d\phi^{1}\right)
     \\
  &  \\
  & \hspace{.5cm}
      +\tfrac{1}{2}(\phi^{1})^{2}
    \left(\imath_{a}\mathcal{D}\phi^{2}\star \mathcal{D}\phi^{2}
    +\mathcal{D}\phi^{2}\wedge \imath_{a}\star \mathcal{D}\phi^{2}\right)
     \\
  &  \\
  & \hspace{.5cm}
    +\tfrac{1}{2}
        \left(\imath_{a}F\wedge \star F
    -F\wedge \imath_{a}\star F\right)
    -\imath_{a}\star V\,.   
  \end{aligned}
\end{equation}

\noindent
Observe that dual fields $\tilde{A},B$ occur in the energy-momentum tensor
through the field strengths $\mathcal{D}\phi^{2}$ and $F$.

The additional terms do not involve the scalars, either, and, therefore

\begin{subequations}
  \begin{align}
    \mathbf{E}_{1}
    & =
      -d\star d\phi^{1}
      +\phi^{1}\mathcal{D}\phi^{2}\wedge \star \mathcal{D}\phi^{2}
      +\star \frac{\partial V}{\partial\phi^{1}}\,,
    \\
    & \nonumber \\
    \mathbf{E}_{2}
    & =
  - d\star j\,,
  \end{align}
\end{subequations}

Furthermore, they do not involve $A$ and, therefore,

\begin{equation}
  \label{eq:EA}
  \mathbf{E}_{A}
  =
  -d\star F +\vartheta\star j\,.
\end{equation}

Now, let us consider the equations of motion of the dual fields which
give duality relations.  The equation of motion of $\tilde{A}$,

\begin{equation}
  \label{eq:EtildeA}
  \mathbf{E}_{\tilde{A}}
  =
  \tilde{\vartheta} \star j -d\left(\tilde{\vartheta} B\right)\,,
\end{equation}

\noindent
gives the duality relation between $\phi^{2}$ (the current $j$) and
$B$ (its field strength $H$) on-shell, when $d\tilde{\vartheta}=0$.

The equation of motion of $B$,

\begin{equation}
  \label{eq:EB}
  \mathbf{E}_{B}
  =
  -\tilde{\vartheta} \left(\star F-\tilde{F} \right)\,,
\end{equation}

\noindent
is the duality relation between $\tilde{A}$ and $A$.

The equations of motion of the components of the embedding tensor are

\begin{equation}
  \label{eq:EvarthetaM}
    \mathbf{E}_{\vartheta_{M}}
    =
      -G^{M}+\star \frac{\partial V}{\partial \vartheta_{M}}\, .
\end{equation}

\noindent
On-shell these equations are the duality relations between the components of the embedding tensor
$\vartheta_{M}$ and the 3-forms $C^{M}$ as given in \cite{Bergshoeff:2009ph}

\begin{equation}
  G^{M}
  =
  \star \frac{\partial V}{\partial\vartheta_{M}}\,.
\end{equation}

In the framework of this theory, these duality relations are only
non-trivial when the corresponding component of the embedding tensor
occurs in the scalar potential. However, it is clear that if those
parameters also occur as coefficients of terms of higher order in the
Riemann curvature, the duality relations will be non-trivial as well.

Once the duality relations implied by the equations of motion of the dual
fields $\tilde{A},B,C^{M}$ and the embedding tensor $\vartheta_{M}$ are taken
into account, the action we are studying describes a very simple model of a
vector field and two scalars, one of which is charged with respect to the
vector field, its dual or a combination of both, coupled to gravity. The use
of the dual vector field as a gauge field is, perhaps, unusual, and demands the
presence of the 2-form $B$, but we can always eliminate this aspect of the
model by setting $\tilde{\vartheta}=0$.

Finally, $\mathbf{\Theta}$ receives contributions from the variations of
$e^{a},\phi^{1},\phi^{2},A^{M},\vartheta_{M}$ but not from those of $B$ or
$C^{M}$, which occur in the action with no derivatives:

\begin{equation}
  \label{eq:Thetaexample}
  \begin{aligned}
    \mathbf{\Theta}(\varphi,\delta\varphi)
    & =
    -\star (e^{a}\wedge e^{b})\wedge \delta \omega_{ab}
    +\star d\phi^{1}\delta\phi^{1}
    +\star j \delta\phi^{2} +\star F\wedge \delta A
    \\
    & \\
    & \hspace{.5cm}
    +\tilde{\vartheta} B\wedge \delta\tilde{A} +C^{M}\delta\vartheta_{M}\,.
  \end{aligned}
\end{equation}

As we have mentioned, the action is invariant under gauge transformations up
to a total derivative that takes the form

\begin{equation}
  \label{eq:totalderivative}
  \delta_{\rm gauge}S
  =
  \int d\left\{\tilde{\vartheta} \Lambda \wedge d\tilde{A}
    +\vartheta_{M}d\chi^{M}
  \right\}\,.
\end{equation}

This total derivative is only defined up to total derivatives and we can make
use of this freedom to obtain gauge-invariant results, if need be.

\subsection{Gauge conserved charges}

We are going to study the effect of all the independent gauge transformations
simultaneously. We will denote all of them by $\delta_{\rm g}$. From the
general variation of the action Eq.~(\ref{eq:generalvariation}) we get

\begin{equation}
  \delta_{\rm g} S
  =
  \int \left\{
    \mathbf{E}_{2} \delta_{\rm g}\phi^{2}
    +\mathbf{E}_{A^{M}}\wedge \delta_{\rm g} A^{M}
    +\mathbf{E}_{B}\wedge \delta_{\rm g} B
    +\mathbf{E}_{C^{M}}\wedge \delta_{\rm g} C^{M}
     +d\mathbf{\Theta}(\varphi,\delta_{\rm g}\varphi)
  \right\}\,,
\end{equation}

\noindent
with

\begin{equation}
  \mathbf{\Theta}(\varphi,\delta_{\rm g}\varphi)
  =
  \star j \delta_{\rm g}\phi^{2} +\star F\wedge \delta_{\rm g} A
  +\tilde{\vartheta} B\wedge \delta_{\rm g}\tilde{A}\,.
\end{equation}

Substituting the above $\delta_{\rm g}$ variations and the expressions
for the equations of
motion and operating, we arrive at


\begin{equation}
  \delta_{\rm g}S
  =
  \int d\mathbf{\Theta}'(\varphi,\delta_{\rm g}\varphi)\,,
\end{equation}

\noindent
with

\begin{equation}
  \begin{aligned}
    \mathbf{\Theta}'(\varphi,\delta_{\rm g}\varphi)
    & =
    \mathbf{\Theta}(\varphi,\delta_{\rm g}\varphi)
        -\star j \vartheta_{M}\sigma^{M} +\sigma d\star F
        +\tilde{\sigma}d\left(\tilde{\vartheta}B\right)
        -\tilde{\vartheta} \left(\star F-\tilde{F} \right)\wedge
        \Lambda
        -d\vartheta_{M}\wedge \chi^{M}
        \\
        & \\
        & =
           d\left(\sigma \star F+\tilde{\sigma}\tilde{\vartheta} B\right)
            +\tilde{\vartheta}d\tilde{A}\wedge\Lambda
        -d\vartheta_{M}\wedge \chi^{M}\,.
  \end{aligned}
\end{equation}

On the other hand, the action is only gauge invariant up to the total derivative
in Eq.~(\ref{eq:totalderivative}) that we can write, for the sake of
convenience, in the form

\begin{equation}
  \delta_{\rm g}S
  =
  \int d\left[
    \tilde{\vartheta}  d\tilde{A} \wedge \Lambda 
    +\vartheta_{M}d\chi^{M} -d\left(\tilde{\vartheta} \tilde{A} \wedge \Lambda\right)
  \right]
  =
  \int d\left[
   \tilde{A} \wedge d\left(\tilde{\vartheta}  \Lambda\right) 
    +\vartheta_{M}d\chi^{M} 
  \right]
  \,,
\end{equation}

\noindent
and combining this result with the previous one we arrive at the off-shell identity

\begin{equation}
  \int d\mathbf{J}
  =
  0\,,
  \,\,\,\,\,
  \text{with}
  \,\,\,\,\,
  \mathbf{J}
  \equiv
  \mathbf{\Theta}'(\varphi,\delta_{\rm g}\varphi)
  -\tilde{A} \wedge d\left(\tilde{\vartheta}  \Lambda\right) 
    -\vartheta_{M}d\chi^{M}   \,,
\end{equation}

\noindent
which implies, locally

\begin{equation}
  \mathbf{J}
  =
  d\mathbf{Q}\,,
\end{equation}

\noindent
where 

\begin{equation}
  \label{eq:2formcharge}
  \mathbf{Q}
  =
  \sigma \star F+\tilde{\sigma}\tilde{\vartheta} B
  +\tilde{\vartheta}\tilde{A}\wedge \Lambda
-\vartheta_{M}\chi^{M}\,.
\end{equation}

Now we must identify the Killing parameters $\sigma^{M},\Lambda,\chi^{M}$ that
generate transformations that leave invariant all the fields.

\label{page:discussion} The 2-form $B$ is left invariant by 1-forms which are
closed, so $\Lambda=h+d\alpha$, for a harmonic 1-form $h$ and an arbitrary
function $\alpha$. However, the 1-forms $A^{M}$ are invariant for functions
$\sigma^{M}$ such that $d\sigma^{M}=\vartheta^{M}(h+d\alpha)$, which implies
that $h=0$ and $\sigma^{M}=\vartheta^{M}\alpha +\beta^{M}$, for arbitrary,
constant symplectic vectors $\beta^{M}$.  The invariance of $\phi^{2}$ implies
that $\vartheta_{M}\beta^{M}=0$. If $\vartheta_{M}\neq 0$, then
$\beta^{M}=\vartheta^{M}\beta$ for a single arbitrary constant $\beta$, but
when $\vartheta_{M}=0$, $\beta^{M}$ is arbitrary. Finally, the invariance of
$C^{M}$ implies that the 2-forms
$\chi^{M}= \vartheta^{M}\alpha+\beta^{M} B+ \Omega^{M}+d\epsilon^{M}$ where
$\epsilon^{M}$ and $\Omega^{M}$ are, respectively, symplectic vectors of
1-forms and harmonic 2-forms. The (pullback of the) latter are proportional to
the volume of the 2-dimensional space on which the charge 2-form is going to
be integrated and we can write, with a slight abuse of language
$\omega^{M}=\gamma^{M}\Omega_{\partial V}$.  Summarizing:

\begin{subequations}
  \label{eq:Killingparametersexample}
  \begin{align}
    \sigma^{M}
    & =
      \vartheta^{M}\alpha+\beta^{M}\,,
    \\
    & \nonumber \\
    \Lambda
    & =
      d\alpha\,,
    \\
    & \nonumber \\
    \chi^{M}
    & =
      \vartheta^{M}\alpha+\beta^{M} B+ \gamma^{M}\Omega_{\partial V}+d\epsilon^{M}\,,
  \end{align}
\end{subequations}

\noindent
with

\begin{equation}
  \vartheta_{M}\beta^{M}
  =
  0\,,
  \hspace{1cm}
  d\star \Omega^{M}
  =
  0\,.
\end{equation}

When the fields are on-shell, these Killing parameters give rise to several
independent conserved charges in the 3-volume $V$ with compact boundary
$\partial V$:

\begin{description}
\item[Electric charge:] associated to $\beta$, which can set to
  1:\footnote{This is the upper component of $\beta^{M}$.}

\begin{equation}
  \label{eq:electricchargedef}
  Q
  \equiv
  \frac{-1}{16\pi G_{N}^{(4)}}
  \int_{\partial V} \left(\star F-\vartheta B \right)\,,
\end{equation}

\noindent
If we deform $\partial V$ without crossing any sources (\textit{i.e.}~points at
which the equations of motion are not satisfied.), the difference between the
charges will be, via Stokes theorem, the volume integral

\begin{equation}
  \Delta \mathbf{Q}
  =
  \frac{-1}{16\pi G_{N}^{(4)}}
  \int_{V} d\left(\star F-\vartheta B \right)  \,,
\end{equation}

\noindent
whose integrand vanishes on-shell.
 
\item[2-form charge:] associated to the function $\alpha$ 

\begin{equation}
  Q[\alpha]
    \equiv
  \frac{-1}{16\pi G_{N}^{(4)}}
  \int_{\partial V}
\left[ \alpha\left( \star F-\vartheta B\right)
  + d\alpha\wedge \tilde{A}
\right]
=
  \frac{1}{16\pi G_{N}^{(4)}}
  \int_{\partial V}
d\left(\vartheta\alpha \tilde{A}\right)=0\,.
\end{equation}

\item[3-form charge:] associated to the space which we are going to
  integrate over\footnote{There are no non-trivial charges associated to the
    1-forms $\epsilon^{M}$ because the integrand is, again, a total
    derivative. We have normalized $ \vartheta_{M}\gamma^{M}=1$}, it is just
  its volume (surface)

\begin{equation}
  Q[V]
    \equiv
  \int_{\partial V}
 \Omega_{\partial V}\,.
\end{equation}

\end{description}

We can also define a magnetic charge

\begin{equation}
  \label{eq:magneticchargedef}
  P
  \equiv
  \frac{-1}{16\pi G_{N}^{(4)}}
  \int_{\partial V} \left(F+\tilde{\vartheta} B \right)\,,
\end{equation}

\noindent
which is conserved in the same sense as the electric one thanks to the Bianchi
identity instead of the equations of motion. This charge can be combined with
the electric one in a symplectic vector

\begin{equation}
  \label{eq:symplecticvectorofcharges}
  \left(
    \begin{array}{c}
      P\\ Q \\
    \end{array}
  \right)
  =
  \left(Q^{M}\right)\,,
  \hspace{1cm}
  Q^{M}
  =
  \frac{-1}{16\pi G_{N}^{(4)}}
  \int_{\partial V} \left(F^{M}-\vartheta^{M} B \right)\,.
\end{equation}

\subsection{The Noether-Wald  charge}

\subsubsection{Transformations of the fields}

As usual, we want to define transformations $\delta_{\xi}$ that annihilate all
the fields of a given solution in a gauge-invariant way for certain parameters
$\xi=k$ which are, in particular, Killing vectors. We have to combine standard
Lie derivatives and $k$-dependent (``compensating'') gauge transformations
into gauge-covariant Lie derivatives.

It is convenient to start by analyzing the 2-form $B$ through its
gauge-invariant 3-form field strength $H=dB$. Due to the Bianchi identity
$dH=0$

\begin{equation}
\delta_{\xi}H = -d\imath_{\xi}H\,.  
\end{equation}

\noindent
When $\xi=k$, there must exist a momentum map 1-form $\mathbf{P}_{k}$ such that 

\begin{equation}
d\mathbf{P}_{k}= -\imath_{k}H\,.  
\end{equation}

Now, the transformation of $B$ is the Lie derivative plus a gauge
transformation with a $\xi$-dependent 1-form parameter $\Lambda_{\xi}$
\begin{equation}
\delta_{\xi} B = -d\imath_{\xi}B-\imath_{\xi}H +d\Lambda_{\xi}\,.
\end{equation}

When $\xi=k$ we can use the definition of the momentum map 1-form $\mathbf{P}_{k}$ to
get

\begin{equation}
d\left(\imath_{k}B-\mathbf{P}_{k}-\Lambda_{k}\right) =0  
\end{equation}

\noindent
which is solved by the choice

\begin{equation}
  \Lambda_{k}
  =
  \imath_{k}B-\mathbf{P}_{k}\,.
\end{equation}

\noindent
Then, we define

\begin{equation}
\delta_{\xi} B = -\left(\imath_{\xi}H +d\mathbf{P}_{\xi}\right)\,.
\end{equation}

\noindent
where the 1-form $\mathbf{P}_{\xi}$ is the momentum map 1-form $\mathbf{P}_{k}$ when
$\xi=k$. With these definitions, $\delta_{k}B=0$ automatically and in a
gauge-invariant fashion.

Let us now consider the gauge-invariant 2-form field strengths $F^{M}$:

\begin{equation}
  \delta_{\xi}F^{M}
  =
  -d\imath_{\xi}F^{M} -\imath_{\xi}dF^{M}
  =
  -d\imath_{\xi}F^{M}
  -\imath_{\xi}\left(d\vartheta^{M}\wedge B+\vartheta^{M}H\right)\,.
\end{equation}

\noindent
On-shell and for $\xi=k$

\begin{equation}
  \delta_{k}F^{M}
  =
  -d\imath_{k}F^{M}
  -\vartheta^{M}\imath_{k}H
  =
  -d\left(\imath_{k}F^{M}-\vartheta^{M}\mathbf{P}_{k}\right)
  =
  0\,,
\end{equation}

\noindent
upon use of the definition of $\mathbf{P}_{k}$. Then, locally, there must exist
momentum maps $P^{M}_{k}$ such that

\begin{equation}
  \imath_{k}F^{M}-\vartheta^{M}\mathbf{P}_{k}
  =
  -dP^{M}{}_{k}\,.
\end{equation}

The transformation of the 1-forms $A^{M}$ is (minus) their Lie derivative plus
a gauge transformation with a $\xi$-dependent 1-form parameter $\Lambda_{\xi}$
which has to  be the same we determined before and a gauge transformation with
$\xi$-dependent 0-form parameters $\sigma^{M}_{\xi}$

\begin{equation}
  \begin{aligned}
    \delta_{\xi}A^{M}
    & =
    -d\imath_{\xi}A^{M} -\imath_{\xi}dA^{M}
    +d\sigma^{M}_{\xi}-\vartheta^{M}\Lambda_{\xi}
    \\
    & \\
    & =
    -d\imath_{\xi}A^{M} -\imath_{\xi}F^{M}
    +d\sigma^{M}_{\xi}+\vartheta^{M}\mathbf{P}_{\xi}\,.
  \end{aligned}
\end{equation}

\noindent
When $\xi=k$

\begin{equation}
  \begin{aligned}
    \delta_{k}A^{M}
    & =
    -d\left(\imath_{k}A^{M} -P^{M}_{k}-\sigma^{M}_{k}\right)
    =0\,,
  \end{aligned}
\end{equation}

\noindent
which is solved by the choice

\begin{equation}
  \sigma^{M}_{k}
  =
  \imath_{k}A^{M} -P^{M}_{k}\,.
\end{equation}

\noindent
Therefore, we define

\begin{equation}
  \label{eq:sigmaMxidef}
  \sigma^{M}_{\xi}
  \equiv 
  \imath_{\xi}A^{M} -P^{M}_{\xi}\,,
\end{equation}

\noindent
and

\begin{equation}
    \delta_{\xi}A^{M}
     =
    -\left(\imath_{\xi}F^{M} +dP^{M}_{\xi}-\vartheta^{M}\mathbf{P}_{\xi}\right)\,.
\end{equation}

\noindent
where, when $\xi=k$, $P^{M}_{\xi}$ and $\mathbf{P}_{\xi}$ are, respectively, the
momentum map 0- and 1-forms. Again, $\delta_{k}A^{M}=0$ automatically and in a
gauge-invariant form .

$\phi^{1}$ is a scalar, and transforms in the standard way

\begin{equation}
  \delta_{\xi}\phi^{1}
  =
  -\pounds_{\xi}\phi^{1}
  =
  -\imath_{\xi}d\phi^{1}\,.
\end{equation}

\noindent
This transformation is assumed to vanish for $\xi=k$. 

The scalar $\phi^{2}$ is, however, a gauge field. It is convenient to analyze,
first, its covariant derivative, which is, actually, gauge-invariant.

\begin{equation}
  \delta_{\xi}\mathcal{D}\phi^{2}
  =
  -d\imath_{\xi}\mathcal{D}\phi^{2}
  -\imath_{\xi} d\mathcal{D}\phi^{2}
  =
  -d\imath_{\xi}\mathcal{D}\phi^{2}
  +\vartheta_{M}\imath_{\xi} F^{M}\,.
\end{equation}

\noindent
On-shell and for $\xi=k$ the following identity must hold

\begin{equation}
  \label{eq:phi2identity}
  -d\left(\imath_{k}\mathcal{D}\phi^{2}+\vartheta_{M}P^{M}_{k}\right)
  =
  0\,,
  \,\,\,\,\,\,
  \Rightarrow
  \,\,\,\,\,\,
  \imath_{k}\mathcal{D}\phi^{2}
  =
  -\vartheta_{M}P^{M}_{k}\,,
\end{equation}

\noindent
where $\sigma^{M}_{\xi}$ defined in Eq.~(\ref{eq:sigmaMxidef}). The
transformation of  $\phi^{2}$ is a combination of (minus) the Lie derivative
and a gauge transformation with parameter $\sigma^{M}_{\xi}$

\begin{equation}
  \delta_{\xi}\phi^{2}
  =
  -\imath_{\xi}d\phi^{2}
  +\vartheta_{M}\sigma^{M}_{\xi}
  =
  -\imath_{\xi}\mathcal{D}\phi^{2}-\vartheta_{M}P^{M}_{\xi}\,,
\end{equation}

\noindent
and $\delta_{k}\phi^{2}$ vanishes identically by virtue of
Eq.~(\ref{eq:phi2identity}).

Let us consider, finally, the 3-forms. As usual, it is convenient to study
their 4-form field strengths first. They are gauge-invariant on-shell only. By
assumption, and because these are 4-forms in 4 dimensions

\begin{equation}
  0=\delta_{k}G^{M}
  =
  -d\imath_{k}G^{M}\,,
  \,\,\,\,\,
  \Rightarrow
  dP^{M}_{G\,k} = -\imath_{k}G^{M}\,,
\end{equation}

\noindent
defining the momentum map 2-forms $P^{M}_{G\,k}$. The transformations of the
3-forms $C^{M}$ must be a combination of their Lie derivatives and gauge
transformations with the parameters $\sigma^{M}_{\xi},\Lambda_{\xi}$ that we
have already determined and, possibly $\chi^{M}_{\xi}$:

\begin{equation}
  \begin{aligned}
    \delta_{\xi} C^{M}
    & =
    -d\imath_{\xi}C^{M} -\imath_{\xi}dC^{M} +d\chi^{M}_{\xi}
    -\sigma^{M}_{\xi}\star j -\vartheta^{M}\Lambda_{\xi}\wedge B
    -\delta^{M\,\tilde{.}}\Lambda_{\xi}\wedge (\star F -\tilde{F})
    \\
    & \\
    & =
    -d\left(\imath_{\xi}C^{M}-\chi^{M}_{\xi}\right)
    -\imath_{\xi}G^{M} +P^{M}_{\xi}\star j
  -A^{M}\wedge \imath_{\xi}\star j
    \\
    & \\
    & \hspace{.5cm}
    +\vartheta^{M}\mathbf{P}_{\xi}\wedge B
    +\delta^{M\,\tilde{.}}\left[\mathbf{P}_{\xi}\wedge(\star F -\tilde{F})
      +B\wedge \imath_{\xi} (\star F -\tilde{F})\right]\,.
  \end{aligned}
\end{equation}

On-shell and for $\xi=k$

\begin{equation}
  \begin{aligned}
    \delta_{k} C^{M}
    & =
    -d\left(\imath_{k}C^{M}-\chi^{M}_{k}-P^{M}_{G\,k}-\mathbf{P}_{k}\wedge A^{M}\right)
+P^{M}_{k}H+\mathbf{P}_{k}\wedge F^{M}\,.
  \end{aligned}
\end{equation}

We can show that the last two terms are, locally, total derivative:

\begin{equation}
  \begin{aligned}
    d\left(P^{M}_{k}H+\mathbf{P}_{k}\wedge F^{M}\right)
    & =
    dP^{M}_{k}H +d\mathbf{P}_{k}\wedge F^{M}-\vartheta^{M}\mathbf{P}_{k}\wedge H
    \\
    & \\
    & =
    -\imath_{k}F\wedge H -\imath_{k}H\wedge F^{M}
=
    -\imath_{k}\left(F\wedge H\right)
=
    0\,.
  \end{aligned}
\end{equation}

\noindent
Thus, we define the  2-form $X^{M}_{2\,k}$ by

\begin{equation}
P^{M}_{k}H+\mathbf{P}_{k}\wedge F^{M}  
\equiv
dX^{M}_{2\,k}\,.
\end{equation}

\noindent
Absorbing it in the definition of $P^{M}_{G\,k}$, which now satisfies

\begin{equation}
  \label{eq:PMGKdef}
  dP^{M}_{G\,k}
  =
  -\imath_{k}G^{M}+P^{M}_{k}H+\mathbf{P}_{k}\wedge F^{M}\,.  
\end{equation}

\noindent
we conclude that

\begin{equation}
    \delta_{k} C^{M}
     =
     -d\left(\imath_{k}C^{M}-\chi^{M}_{k}-P^{M}_{G\,k} -\mathbf{P}_{k}\wedge A^{M}\right)
    =
    0\,,
\end{equation}

\noindent
which is solved by

\begin{equation}
  \chi^{M}_{k}
  =
  \imath_{k}C^{M}-P^{M}_{G\,k}-\mathbf{P}_{k}\wedge A^{M}\,.
\end{equation}

\noindent
Then, we arrive at the definition

\begin{equation}
  \begin{aligned}
        \delta_{\xi} C^{M}
& =
-\imath_{\xi}G^{M}-dP^{M}_{G\,\xi}+\mathbf{P}_{\xi}\wedge F^{M} +P^{M}_{\xi}\star j
  -A^{M}\wedge \left(\imath_{\xi}\star j+d\mathbf{P}_{\xi}\right)
    \\
    & \\
    & \hspace{.5cm}
    +\delta^{M\,\tilde{.}}\left[\mathbf{P}_{\xi}\wedge(\star F -\tilde{F})
      +B\wedge \imath_{\xi} (\star F -\tilde{F})\right]\,,
  \end{aligned}
\end{equation}

\noindent
which vanishes automatically for $\xi=k$.

Summarizing, the transformations that we are going to consider are

\begin{subequations}
  \begin{align}
    \delta_{\xi}e^{a}
    & =
      -(\mathcal{D}\xi^{a}+P_{\xi}{}^{a}{}_{b}e^{b})\,,
    \\
    & \nonumber \\
    \delta_{\xi}\omega^{ab}
    & =
      -(\imath_{\xi}R^{ab}+\mathcal{D}P_{\xi}{}^{ab})\,,
    \\
    & \nonumber \\
      \delta_{\xi}\phi^{1}
    & =
      -\imath_{\xi}d\phi^{1}\,,
    \\
    & \nonumber \\
  \delta_{\xi}\phi^{2}
  & =
  -\left(\imath_{\xi}\mathcal{D}\phi^{2} +\vartheta_{M}P^{M}_{\xi}\right)\,,
\\
    & \nonumber \\
        \delta_{\xi}A^{M}
     & =
       -\left(\imath_{\xi}F^{M} +dP^{M}_{\xi}-\vartheta^{M}\mathbf{P}_{\xi}\right)\,,
    \\
    & \nonumber \\
    \delta_{\xi} B
    & =
      -\left(\imath_{\xi}H +d\mathbf{P}_{\xi}\right)\,,
    \\
    & \nonumber \\
        \delta_{\xi} C^{M}
& =
-\left(\imath_{\xi}G^{M}+dP^{M}_{G\,\xi}-\mathbf{P}_{\xi}\wedge F^{M} -P^{M}_{\xi}\star j\right)
  -A^{M}\wedge \left(\imath_{\xi}\star j+d\mathbf{P}_{\xi}\right)
    \nonumber \\
    & \nonumber \\
    & \hspace{.5cm}
    +\delta^{M\,\tilde{.}}\left[\mathbf{P}_{\xi}\wedge(\star F -\tilde{F})
      +B\wedge \imath_{\xi} (\star F -\tilde{F})\right]\,,
    \\
    & \nonumber \\
    \delta_{\xi} \vartheta_{M}
    & =
      -\imath_{\xi}d\vartheta_{M}\,,
  \end{align}
\end{subequations}

\noindent
and the momentum maps 0-, 1-, and 2-forms satisfy

\begin{subequations}
  \begin{align}
  \mathcal{D}P_{k}{}^{ab}
 & =
    -\imath_{k}R^{ab}\,,
    \\
    & \nonumber \\
    dP^{M}{}_{k}
    & =
      -\imath_{k}F^{M}+\vartheta^{M}\mathbf{P}_{k}\,,
    \\
    & \nonumber \\
    d\mathbf{P}_{k}
    & =
      -\imath_{k}H\,,
    \\
    & \nonumber \\
    \label{eq:PGkequation}
  dP^{M}_{G\,k}
  & =
  -\imath_{k}G^{M}+P^{M}_{k}H+\mathbf{P}_{k}\wedge F^{M}\,.  
  \end{align}
\end{subequations}

Furthermore, when $\xi=k$

\begin{equation}
    \imath_{k}\mathcal{D}\phi^{2}
     =
      -\vartheta_{M}P^{M}_{k}\,.
\end{equation}

\subsubsection{Transformation of the action}

Substituting the above transformations of the fields in

\begin{equation}
  \label{eq:deltaxiSexample}
  \begin{aligned}
  \delta_{\xi} S
 & =
  \int \left\{
    \mathbf{E}_{a}\wedge  \delta_{\xi} e^{a}
    +\mathbf{E}_{1} \delta_{\xi}\phi^{1}
    +\mathbf{E}_{2}  \delta_{\xi}\phi^{2}
    +\mathbf{E}_{A^{M}}\wedge  \delta_{\xi} A^{M}
    +\mathbf{E}_{B}\wedge  \delta_{\xi} B
    +\mathbf{E}_{C^{M}}\wedge  \delta_{\xi} C^{M}
  \right.
  \\
  & \\
  & \hspace{.5cm}
  \left.
        +\mathbf{E}_{\vartheta_{M}}\wedge \delta_{\xi} \vartheta_{M}
     +d\mathbf{\Theta}(\varphi, \delta_{\xi}\varphi)
  \right\}\,,
\end{aligned}
\end{equation}
integrating by parts 
and using the Noether identities we are left with

\begin{equation}
  \label{eq:deltaxiSexample4}
  \delta_{\xi} S
 =
  \int d\mathbf{\Theta}'(\varphi,\delta_{\xi}\varphi)\,,
\end{equation}

\noindent
with

\begin{equation}
  \label{eq:Thetaprimeexample}
    \mathbf{\Theta}'(\varphi,\delta_{\xi}\varphi)
    \equiv
    \mathbf{\Theta}(\varphi,\delta_{\xi}\varphi) +\xi^{a}\mathbf{E}_{a}
    +P^{M}_{\xi}\mathbf{E}_{A^{M}}
    -\mathbf{P}_{\xi}\wedge\left(\mathbf{E}_{B} +\mathbf{E}_{C^{M}}\wedge A^{M}\right)
    +P^{M}_{G\,\xi}\wedge \mathbf{E}_{C^{M}}\,.  
\end{equation}

Under these transformations, the action transforms into the integral of a
total derivative, that we have chosen so as to obtain a final gauge-invariant
result:

\begin{equation}
  \label{eq:otramasdeltaxiS}
  \delta_{\xi} S
  =
  \int d\left\{-\imath_{\xi}\mathbf{L}
    +\tilde{\vartheta}\imath_{\xi}B \wedge d\tilde{A}
      +\tilde{A}\wedge d\left(\tilde{\vartheta}\mathbf{P}_{\xi}\right)
    -d\vartheta_{M}\wedge \left(\imath_{\xi}C^{M}-\mathbf{P}_{\xi}\wedge A^{M}\right)
    -\vartheta_{M}dP^{M}_{G\,\xi}\right\}\,.
\end{equation}

\noindent
Equating this result for $\delta_{\xi} S$ with the one in
Eq.~(\ref{eq:deltaxiSexample4}) we arrive to the identity

\begin{equation}
  \int d\mathbf{J}[\xi]=0\,,
\end{equation}

\noindent
with

\begin{equation}
  \label{eq:Jxidefexample}
    \mathbf{J}[\xi]
    =
    \mathbf{\Theta}'(\varphi,\delta_{\xi}\varphi)
+\imath_{\xi}\mathbf{L}
    -\tilde{\vartheta}\imath_{\xi}B \wedge d\tilde{A}
      -\tilde{A}\wedge d\left(\tilde{\vartheta}\mathbf{P}_{\xi}\right)
    +d\vartheta_{M}\wedge \left(\imath_{\xi}C^{M}-\mathbf{P}_{\xi}\wedge A^{M}\right)
    +\vartheta_{M}dP^{M}_{G\,\xi}\,.
\end{equation}

Simplifying this expression  we get

\begin{subequations}
  \begin{align}
    \mathbf{J}[\xi]
    & =
    d\mathbf{Q}[\xi]\,,
    \\
    & \nonumber \\
    \label{eq:Qxiexample}
    \mathbf{Q}[\xi]
    & =
    \star (e^{a}\wedge e^{b})\wedge P_{\xi\, ab}
      -\left(P_{\xi}\star F +\tilde{\vartheta}\tilde{P}_{\xi} B
      +\tilde{\vartheta}\tilde{A}\wedge \mathbf{P}_{\xi}
      -\vartheta_{M}P^{M}_{G\,\xi} \right)\,.
  \end{align}
\end{subequations}

The second term in this formula, in parenthesis, should be compared with the
2-form charge associated to gauge transformations Eq.~(\ref{eq:2formcharge}).

\subsection{Generalized, restricted, zeroth laws}
\label{sec-zerothlawexample}

We just need to adapt the discussion in Section~\ref{sec-zerothlaw} to the
model at hand, which has more fields. On the bifurcation surface
$\mathcal{BH}$ we have

\begin{subequations}
  \begin{align}
    dP^{M}_{k} -\vartheta^{M}\mathbf{P}_{k}
    & \stackrel{\mathcal{BH}}{=}
      0\,,
    \\
    & \nonumber \\
    d\mathbf{P}_{k}
    & \stackrel{\mathcal{BH}}{=}
      0\,,
    \\
    & \nonumber \\
    d\mathbf{P}^{M}_{G\, k} -P^{M}_{k}H -\mathbf{P}_{k}\wedge F^{M}
    & \stackrel{\mathcal{BH}}{=}
      0\,,
    \\
    & \nonumber \\
    \label{eq:ThetaP-Ortogonal}
    \vartheta_{M}P^{M}_{k}
    & \stackrel{\mathcal{BH}}{=}
      0\,.    
  \end{align}
\end{subequations}

These equations are equivalent to the equations that the Killing parameters
discussed on page~\ref{page:discussion} must satisfy: first of all, the second
equation implies that $\mathbf{P}_{k}=h+d\alpha$, where $h$ is a harmonic
1-form on the bifurcation surface and $\alpha$ and arbitrary function. However,
the first equation tells us that $h$ has to be removed from that identity and
$P^{M}_{k}=\vartheta^{M}\alpha+\beta^{M}$ for $\beta^{M}$ which is constant
over the bifurcation surface. The last equation implies that
$\beta^{M}=\vartheta^{M}\beta$ if $\vartheta_{M}\neq 0$, but it is arbitrary
when $\vartheta_{M}= 0$. The third equation takes the form

\begin{equation}
  d\mathbf{P}^{M}_{G\, k} -\left(\vartheta^{M}\alpha+\beta^{M}\right) dB
  -d\alpha \wedge F^{M}
  =
  d\left(\mathbf{P}^{M}_{G\, k} -\alpha F^{M}
    -\beta^{M} B
\right)
     \stackrel{\mathcal{BH}}{=}
      0\,,
\end{equation}
 
\noindent
and, summarizing, we have\footnote{Compare with
  Eqs.~(\ref{eq:Killingparametersexample}).}

\begin{subequations}
  \label{eq:generalizedrestrictedzerothlaws}
  \begin{align}
    P^{M}_{k}
    & \stackrel{\mathcal{BH}}{=}
      \vartheta^{M}\alpha+\beta^{M}\,,
    \\
    & \nonumber \\
    \mathbf{P}_{k}
    & \stackrel{\mathcal{BH}}{=}
      d\alpha\,,
    \\
    & \nonumber \\
    \mathbf{P}^{M}_{G\, k} 
    & \stackrel{\mathcal{BH}}{=}
      \alpha F^{M}
    +\beta^{M} B+\gamma^{M}\Omega_{\mathcal{BH}}\,,
  \end{align}
\end{subequations}

\noindent
where $\gamma^{M}$ is a constant symplectic vector and $\Omega_{\mathcal{BH}}$
is the volume 2-form of the bifurcation surface; from
Eq.~(\ref{eq:ThetaP-Ortogonal}) we have

\begin{equation}
  \label{eq:betaconstraint}
  \vartheta_{M}\beta^{M}
  =
  0\,.
\end{equation}

The components of the constant vector $\beta^{M}$ can be interpreted as the
electric and magnetic potentials over the bifurcation surface and the fact
that they are constant is the generalized zeroth law restricted to the
bifurcation surface. It is unclear whether this property can be extended to
the whole event horizon in this particularly complex model or, at least, it is
unclear how to prove it. However, we will not need this proof.
As the $\beta^{M}$ are the
thermodynamical potentials associated to the electric and magnetic charges,
we will denote them by

\begin{equation}
  \label{eq:PhiMdef}
  \Phi^{M}\equiv \beta^{M}\,,
  \,\,\,\,\,
  \text{with}
  \,\,\,\,\,
  (\Phi^{M})
  =
  \left(
    \begin{array}{c}
      \Phi \\ \widetilde{\Phi} \\
    \end{array}
  \right)\,,
\end{equation}

Observe that Eq.~(\ref{eq:betaconstraint}) becomes a constraint on these
potentials:

\begin{equation}
  \label{eq:phiconstraint}
  \vartheta_{M}\Phi^{M}
  =
  0\,.
\end{equation}

The components of the constant vector $\gamma^{M}$ are the thermodynamical
potentials (``volumes'') associated to the thermodynamical variables
$\vartheta_{M}$ (``pressures''). Again, in order to make contact with the
conventions of Ref.~\cite{Ortin:2021ade}, we can define the potentials $\Theta^{M}$

\begin{equation}
  \label{eq:ThetaMdef}
  \frac{\gamma^{M}}{16\pi G_{N}^{(4)}}
  \equiv
  -\Theta^{M}/V_{\mathcal{BH}}\,,
\end{equation}

\noindent
where $V_{\mathcal{BH}}$ is the volume of the bifurcation surface

\begin{equation}
  V_{\mathcal{BH}}
  =
  \int_{\mathcal{BH}}\Omega_{\mathcal{BH}}\,.
\end{equation}

The fact that the vector $\Theta^{M}$ is constant over the bifurcation surface is
another generalized, restricted, zeroth law

There is no role for the function $\alpha$: there are no conserved charges
associated to the gauge transformations $\delta_{\Lambda}$ and $\alpha$ will
also drop out of the Smarr formula.

\subsection{Komar integral and Smarr formula}

We are now ready to construct the Komar integral for this theory along the
lines explained in Section~\ref{sec-Komar}. It provides a highly non-trivial
check of the Noether-Wald charge.

Let us consider a field configuration that satisfies the all the equations of
motion and an infinitesimal diffeomorphism $\xi=k$ that generates a symmetry
of the whole field configuration. Then, since
$\mathbf{\Theta}(\varphi,\delta_{\xi}\varphi)$ is linear in
$\delta_{\xi}\varphi$, it vanishes when $\xi=k$ and, since the equations of
motion are satisfied, so does $\mathbf{\Theta}(\varphi,\delta_{\xi}\varphi)$.
Then, from from the definition Eq.~(\ref{eq:Jxidefexample}) we find
that\footnote{As before, we use $\doteq$ for identities that only hold on-shell.}

\begin{equation}
  \label{eq:Jkexample}
    \mathbf{J}[k]
    \,\doteq\,
\imath_{k}\mathbf{L}
    -\tilde{\vartheta}\imath_{\xi}B \wedge d\tilde{A}
      -\tilde{A}\wedge d\left(\tilde{\vartheta}\mathbf{P}_{\xi}\right)
 +d\vartheta_{M}\wedge \left(\imath_{\xi}C^{M}-\mathbf{P}_{\xi}\wedge A^{M}\right)
    +\vartheta_{M}dP^{M}_{G\,\xi}\,.
\end{equation}

On the other hand, by construction,

\begin{equation}
\delta_{k}S=0\,.
\end{equation}

\noindent
and, thus, the total derivative in Eq.~(\ref{eq:otramasdeltaxiS}) evaluated
for $\xi=k$, which coincides with the on-shell value of $ \mathbf{J}[k]$ in
Eq.~(\ref{eq:Jkexample}), must vanish identically and, locally, there is a
2-form $\varpi_{k}$ such that

\begin{equation}
  d\varpi_{k}
  \,\doteq\,
  \imath_{k}\mathbf{L}
    -\tilde{\vartheta}\imath_{\xi}B \wedge d\tilde{A}
      -\tilde{A}\wedge d\left(\tilde{\vartheta}\mathbf{P}_{\xi}\right)
 +d\vartheta_{M}\wedge \left(\imath_{k}C^{M}-\mathbf{P}_{k}\wedge A^{M}\right)
    +\vartheta_{M}dP^{M}_{G\,k}
    =
     \mathbf{J}[k]\,.
\end{equation}

Since we also have $\mathbf{J}[k]= d\mathbf{Q}[k]$, we conclude that

\begin{equation}
  d\left\{ \mathbf{Q}[k]-\varpi_{k}\right\}
  \,\doteq\,
  0\,.
\end{equation}

In order to compute the Komar charge
$ -\left\{\mathbf{Q}[k]-\varpi_{k}\right\}$ we proceed as before, taking the
trace of the Einstein equations (\ref{eq:Eaexample})

\begin{equation}
  \begin{aligned}
    e^{a}\wedge E_{a}
    & =
    -2\mathbf{L} + F\wedge \star F
+2\tilde{\vartheta}B\wedge \left(\tilde{F}-\tfrac{1}{2}\vartheta B\right)
-2C^{M}\wedge d\vartheta_{M}-2\star V \,.   
  \end{aligned}
\end{equation}

\noindent
and, on-shell ($\star F=\tilde{F}$ and $d\vartheta_{M}=0$)



\begin{equation}
    \imath_{k}\mathbf{L}
     \,\doteq\,
  \tfrac{1}{2}\left(\imath_{k}F+2\tilde{\vartheta}\imath_{k}B\right)\wedge \tilde{F}
  +\tfrac{1}{2}\left(F+2\tilde{\vartheta}B\right)\wedge \imath_{k}\tilde{F}
-\tilde{\vartheta}\vartheta \imath_{k}B\wedge B
    -\imath_{k}\star V \,.
\end{equation}

\noindent
Combining this result with the other terms and operating, we get

\begin{equation}
  \begin{aligned}
    \imath_{k}\mathbf{L}
    -\tilde{\vartheta}\imath_{\xi}B \wedge d\tilde{A}
      -\tilde{A}\wedge d\left(\tilde{\vartheta}\mathbf{P}_{\xi}\right)
    +d\vartheta_{M}\wedge \left(\imath_{k}C^{M}-\mathbf{P}_{k}\wedge A^{M}\right)
 +\vartheta_{M}dP^{M}_{G\,k}
 & \,\doteq\,
 \\
& \\
 \,\doteq\,
 \tfrac{1}{2}\vartheta_{M}
 \left( \mathbf{P}_{k}\wedge  dA^{M}-B\wedge dP^{M}_{k} \right)
-\imath_{k}\star V
+d\left(\vartheta_{M}P^{M}_{G\,k}
 -\tfrac{1}{2}P_{k} d\tilde{A}
 -\tfrac{1}{2} \tilde{P}_{k}dA
-\tilde{\vartheta}\tilde{A}\wedge \mathbf{P}_{k}\right)\,.
  \end{aligned}
\end{equation}

Let us now consider the term involving the scalar potential:
using the fact that the potential is a homogeneous function of the
embedding tensor, Eq.~(\ref{eq:PotHomFnctDeg2}), and the
on-shell $\vartheta_{M}$ equation of motion, Eq.~(\ref{eq:EvarthetaM}),
we obtain
%
%

\begin{equation}
  \imath_{k}\star V
  =
  \tfrac{1}{2}\vartheta_{M}\imath_{k}G^{M}
  =
  \tfrac{1}{2}\vartheta_{M}\left(
    -dP^{M}_{G\,k}
  +P^{M}_{k}H+\mathbf{P}_{k}\wedge F^{M}\right)\,.
\end{equation}

\noindent
After use of the definition of the 2-form momentum map $P^{M}_{G\, k}$ in
Eq.~(\ref{eq:PGkequation}), we arrive at

\begin{equation}
  \begin{aligned}
    d\varpi_{k}
    & =
    \imath_{k}\mathbf{L}
    -\tilde{\vartheta}\imath_{\xi}B \wedge d\tilde{A}
      -\tilde{A}\wedge d\left(\tilde{\vartheta}\mathbf{P}_{\xi}\right)
    +d\vartheta_{M}\wedge \left(\imath_{k}C^{M}-\mathbf{P}_{k}\wedge A^{M}\right)
    +\vartheta_{M}dP^{M}_{G\,k}
    \\
    & \\
 & \,\doteq\,
d\left(\tfrac{3}{2}\vartheta_{M}P^{M}_{G\,k}
 -\tfrac{1}{2}P_{k} d\tilde{A}
 -\tfrac{1}{2} \tilde{P}_{k}dA
 -\tfrac{1}{2}\vartheta_{M}P^{M}_{k}B
+\tilde{\vartheta}\tilde{A}\wedge \mathbf{P}_{k}\right)\,.
  \end{aligned}
\end{equation}

Combining this result with Eq.~(\ref{eq:Qxiexample}) we obtain the
Komar charge

\begin{equation}
  \label{eq:Komarcharge}
  -\left\{\mathbf{Q}[k]-\varpi_{k}\right\}
  \,\doteq\,
\frac{-1}{16\pi G_{N}^{(4)}}\left\{ \star (e^{a}\wedge e^{b})\wedge P_{k\, ab}
      +\tfrac{1}{2}P_{k\, M}F^{M}
      -\tfrac{1}{2}\vartheta_{M}P^{M}_{G\,k}
      \right\}\,.
\end{equation}

This is a manifestly (formally) symplectic-invariant result
\cite{Mitsios:2021zrn} that reduces to the result obtained in
Section~\ref{sec-Komar} if we eliminate the scalar and 1-form fields;
it reduces to the Einstein-Maxwell result upon setting the embedding tensor to zero.

We can now proceed as in Section~\ref{sec-Komar} to derive the Smarr formula
through the identity between the Komar integrals over the bifurcation surface
and at spatial infinity Eq.~(\ref{eq:Komaridentity2}). At infinity 

\begin{equation}
  \mathcal{K}(S^{2}_{\infty})
  =
  \frac{-1}{16\pi G_{N}^{(4)}}\int_{S^{2}_{\infty}}
  \left\{ \star (e^{a}\wedge e^{b})\wedge P_{k\, ab}
      +\tfrac{1}{2}P_{k\, M}F^{M}
      -\tfrac{1}{2}\vartheta_{M}P^{M}_{G\,k}
    \right\}
    =
    \tfrac{1}{2}\left(M-\Omega J\right)\,,
\end{equation}

\noindent
essentially by definition. Over the bifurcation surface, using the
generalized, restricted, zeroth laws
Eqs.~(\ref{eq:generalizedrestrictedzerothlaws}), the definitions of the
potentials Eqs.~(\ref{eq:PhiMdef}) (\ref{eq:ThetaMdef}) and of the definitions
of electric and magnetic charges Eq.~(\ref{eq:symplecticvectorofcharges}), we get

\begin{equation}
  \begin{aligned}
    \mathcal{K}(\mathcal{BH})
    & =
    ST
      +\tfrac{1}{2}\Phi_{M}Q^{M}
      -\tfrac{1}{2}\vartheta_{M}\Theta^{M}\,,
    \end{aligned}
\end{equation}

\noindent
and we obtain the Smarr formula

\begin{equation}
  \label{eq:Smarrformulaexample}
M= 2ST +\Omega J +\Phi_{M}Q^{M} -\vartheta_{M}\Theta^{M}\,.  
\end{equation}

In order to check this formula we need explicit analytic solutions of the
equations of motion of this model, but this is quite difficult, as the attempt
made in the appendix shows. However, it is clear that, had we considered an
additional cosmological-constant parameter $\lambda$ in the theory, we would
simply have obtained an additional term $-\lambda\Theta_{\lambda}$ in the
above formula. That formula should remain valid when the embedding tensor is
set to zero, in which case the theory reduces to the cosmological
Einstein-Maxwell theory. Still, it is quite difficult to check the Smarr
formula explicitly because the radius of the horizon is the solution of a
quartic equation, except in particular cases such as for \textit{cold} black
holes, which have extremal, zero temperature horizons \cite{Romans:1991nq}.

Observe that, due to the constraint Eq.~(\ref{eq:phiconstraint}), only one
combination of the electric and magnetic potentials occurs in the above
formula and, henceforth, only one combination of the electric and magnetic
charges does.

\subsection{The first law and black-hole chemistry}

It is not necessary to repeat here all the steps that lead to the first law

\begin{equation}
\delta M = T\delta S +\Omega\delta J +  \Phi \delta Q +\Theta^{M}\delta \vartheta_{M}\,.
\end{equation}

Observe that, as usual, only the variation of the electric charge and its
associated electric potential occur in the first law. This could be due to a
limitation of the techniques that we are using. Nevertheless, if the magnetic
counterpart of the $ \Phi \delta Q$ term was present, due to the constraint
Eq.~(\ref{eq:phiconstraint}), there would be a combination of electric and
magnetic charges the mass of the black whole would be independent of. We will
comment upon this point in the discussion section.

\section{Discussion}
\label{sec-discussion}

In this paper we have shown how the variations of the cosmological constant
and other dimensionful constants occurring in a theory of gravity can be
consistently dealt with and understood in the framework of Wald's formalism and
how they enter the first law of black-hole thermodynamics and the Smarr
formula.  In the example that we have completely worked out in
Section~\ref{sec-generalexample}, the constants that we have considered can be
seen as components of the embedding tensor (a very simple one since there is
only a 1-dimensional symmetry to be gauged) and our result proves the
conjectured role of the embedding tensor as a thermodynamical variable.

A very interesting aspect of the Smarr formula is that, if it is general
enough and it includes all the charges a black hole can carry and all the
moduli of the theory under consideration, then it has to be invariant under
all the duality transformations. Observe that duality transformations act on
the moduli and charges but leave the mass, temperature and entropy invariant
because the Einstein metric is left invariant by them. In
Ref.~\cite{Mitsios:2021zrn} we showed that, in the context of pure
$\mathcal{N}=4,d=4$ supergravity, indeed, the term involving the electric and
magnetic potentials and charges is formally symplectic invariant. This
automatically implies its invariance under the SO$(6)\times$SL$(2,\mathbb{R})$
duality group of $\mathcal{N}=4,d=4$ supergravity since all the 4-dimensional
duality groups act on the 1-form fields as a subgroup of the symplectic group
\cite{Gaillard:1981rj}.  The same happens in the very simple example that we
have considered here but we have also seen that the term involving the
embedding tensor and its conjugate thermodynamical potential is also
electric-magnetic duality invariant as it should, according to the general
arguments given above. In more general models the embedding tensor is denoted
by $\vartheta_{A}{}^{M}$, where the index $A$ runs over the Lie algebra of the
symmetry group of the theory.  The terms that must occur in the first law and
in the Smarr formula must be, respectively, of the form

\begin{equation}
  -\Theta^{A}{}_{M}\delta \vartheta_{A}{}^{M}\,,
  \,\,\,\,\,\,
  \text{and}
  \,\,\,\,\,\,
  +\Theta^{A}{}_{M}\vartheta_{A}{}^{M}\,.
\end{equation}

Thus, in general 4-dimensional theories with an arbitrary number of 1-form
fields labeled by $I$, we expect the first law and the Smarr formula to take
the general form\footnote{$\Phi_{M}Q^{M}=\Phi^{I}Q_{I}-\tilde{\Phi}_{I}P^{I}$}

\begin{subequations}
  \begin{align}
    \delta M
    & =
      T\delta S +\Omega\delta J +  \Phi_{I} \delta Q^{I}
      -\Theta^{A}{}_{M}\delta \vartheta_{A}{}^{M}\,,
    \\
    & \nonumber \\
    M
    & =
      2ST +\Omega J +\Phi_{M}Q^{M} +\Theta^{A}{}_{M}\vartheta_{A}{}^{M}\,.
  \end{align}
\end{subequations}

We expect to verify the validity of this general formula in more general
models of gauge supergravity in forthcoming works.

Concerning the particular model that we have constructed and studied in
Section~\ref{sec-generalexample} to test these ideas, as we pointed out
before, only one combination of the electric and magnetic potentials may occur
in the first law. Therefore, there would be a combination of electric and
magnetic charges the mass of the black holes of this theory would not depend
on. In order to check this quite unusual property it is necessary to find the
most general black-hole solutions of the theory. This is a very complicated
problem.  In the appendix we have managed to find solutions with one charge
(the embedding of the Reissner-Nordstr\"om-(A)DS black hole in this theory) for
a particularly simple choice of embedding tensor, but these solutions are not
general enough to check whether this property, predicted by the first law, that
is true.\footnote{Here we are assuming that the complete first law should
  include a term proportional to the variation of the magnetic charge that we
  still do not know how to incorporate in our formalism. This is a problem on
  which we hope to report in forthcoming work.} Further work in this direction
is necessary and under way.

\section*{Acknowledgments}

This work has been supported in part by the MCIU, AEI, FEDER (UE) grants
PGC2018-095205-B-I00 {\&} PGC2018-096894-B-I00, the Principado de Asturias
grant SV-PA-21-AYUD/2021/52177 and by the Spanish Research Agency (Agencia
Estatal de Investigaci\'on) through the grant IFT Centro de Excelencia Severo
Ochoa CEX2020-001007-S.  Dimitrios Mitsios acknowledges support by the Onassis
Foundation under scholarship ID: F ZR 038/1-2021/2022 and the ERC-SyG project
{\em Recursive and Exact New Quantum Theory} (ReNewQuantum) which received
funding from the European Research Council (ERC) under the European Union's
Horizon 2020 research and innovation programme under grant agreement No
810573.  TO wishes to thank M.M.~Fern\'andez for her permanent support.

\appendix

\section{Searching for solutions}
\label{sec-solutions}

We would like to have a black-hole solution of the theory introduced in
Section~\ref{sec-generalexample} in order to test the general results that we
have derived. For the sake of simplicity, we set $\tilde{\vartheta}=0$
(electric gauging) and we set $\vartheta=g$, constant. We can set $B=0$ (so
$F=dA$)and ignore $C$. The equations of motion that remain to be solved are

\begin{subequations}
  \begin{align}
      \mathbf{E}_{a}
  & =
    \imath_{a}\star (e^{c}\wedge e^{d})\wedge R_{cd}
      +\tfrac{1}{2}
    \left(\imath_{a}d\phi^{1}\star d\phi^{1}
    +d\phi^{1}\wedge \imath_{a}\star d\phi^{1}\right)
\nonumber      \\
  &  \nonumber \\
  & \hspace{.5cm}
      +\tfrac{1}{2}
    \left(\imath_{a}\mathcal{D}\phi^{2}\star j
    +\mathcal{D}\phi^{2}\wedge \imath_{a}\star j \right)
    \nonumber  \\
  &  \nonumber \\
  & \hspace{.5cm}
    +\tfrac{1}{2}
        \left(\imath_{a}F\wedge \star F
    -F\wedge \imath_{a}\star F\right)
    -\imath_{a}\star V\,,
    \\
    & \nonumber \\
        \mathbf{E}_{1}
    & =
      -d\star d\phi^{1}
      +\phi^{1}\mathcal{D}\phi^{2}\wedge \star \mathcal{D}\phi^{2}
      +\star \frac{\partial V}{\partial\phi^{1}}\,,
    \\
    & \nonumber \\
    \mathbf{E}_{2}
    & =
  - d\star j\,,
    \\
    & \nonumber \\
      \mathbf{E}_{A}
  & =
  -d\star F +\vartheta\star j\,,
  \end{align}
\end{subequations}

\noindent
equated to zero.

In the search for solutions, it is convenient to express these equations in
component language:

\begin{subequations}
  \begin{align}
    G_{\mu\nu} +\tfrac{1}{2}\left(\partial_{\mu}\phi^{1}\partial_{\nu}\phi^{1}
    -\tfrac{1}{2}g_{\mu\nu}(\partial\phi^{1})^{2}\right)
    +\tfrac{1}{2}\left(\mathcal{D}_{\mu}\phi^{2}j_{\nu}
    -\tfrac{1}{2}g_{\mu\nu}\mathcal{D}^{\rho}\phi^{2}j_{\rho}\right)
    & \nonumber \\
    & \nonumber \\
    -\tfrac{1}{2}\left(F_{\mu}{}^{\rho}F_{\nu\rho}-\tfrac{1}{4}g_{\mu\nu}F^{2}\right)
    +\tfrac{1}{2}g_{\mu\nu}V
    & =
      0\,,
    \\
    & \nonumber \\
    \label{eq:phi1eq}
    -\nabla^{2}\phi^{1} +\phi^{1}\left(\mathcal{D}\phi^{2}\right)^{2}
    -\partial_{\phi^{1}}V
    & =
      0\,,
    \\
    & \nonumber \\
    \label{eq:phi2eq}
    -\nabla_{\mu}j^{\mu}
    & =
      0\,,
    \\
    & \nonumber \\
    \label{eq:Maxwell}
    \nabla_{\mu}F^{\mu\nu}-gj^{\nu}
    & =
      0\,.
  \end{align}
\end{subequations}

We are interested in static, spherically-symmetric solutions with a metric of
the form

\begin{equation}
  ds^{2}
  =
  \lambda dt^{2} -\lambda^{-1}dr^{2}- R^{2}d\Omega^{2}_{(2)}\,,
\end{equation}

\noindent
where $\lambda$ and $R$ are functions of $r$ to be determined and 

\begin{equation}
  d\Omega^{2}_{(2)}
  =
  d\theta^{2}+\sin^{2}{\theta}d\varphi^{2}\,.
\end{equation}

The timelike Killing vector is $k=\partial_{t}$ and we assume that it
generates a diffeomorphism that leaves invariant all the fields. This means
that

\begin{subequations}
  \begin{align}
    \partial_{t}\phi^{1}
    & =
      0\,,
    \\
    & \nonumber \\
    \mathcal{D}_{t}\phi^{2}
    & =
    \partial_{t}\phi^{2}
      -gA_{t}
      =
      -gP_{k}\,,
    \\
    & \nonumber \\
    F_{t\mu}
    & =
      -\partial_{\mu}P_{k}\,.
  \end{align}
\end{subequations}

If we assume that the electromagnetic field is electric and we work in the
gauge in which the only non-trivial component is $A_{t}$ and it is only a
function of $r$, then the scalars $\phi^{1,2}$ only depend on $r$ as well and

\begin{equation}
  \label{eq:PkAt}
  P_{k}=A_{t}\,.
\end{equation}

The $r$ component of the Maxwell equation (\ref{eq:Maxwell}), tells us
that

\begin{equation}
  j^{r}=0\,,
  \,\,\,\,\,
  \Rightarrow
  \,\,\,\,\,
  \phi^{2}= \text{constant}\,.
\end{equation}

This automatically solves Eq.~(\ref{eq:phi2eq}) and simplifies
Eq.~(\ref{eq:phi1eq}), which can be written in the form upon use of
Eq.~(\ref{eq:PkAt})

\begin{equation}
  \label{eq:phi1eq-2}
  \frac{1}{R^{2}}\left(R^{2}\lambda \phi^{1\,\prime}\right)'
  +g^{2}\phi^{1}\lambda^{-1}P_{k}^{2}-\partial_{\phi^{1}}V
  =
  0\,.
\end{equation}

The $t$ component of the Maxwell equation takes the form

\begin{equation}
  \label{eq:Maxwell2}
  -\frac{1}{R^{2}} \left(R^{2} P_{k}'\right)'
  +g^{2}\lambda^{-1}(\phi^{1})^{2}P_{k}
  =
  0\,.
\end{equation}

Now it is the turn of the Einstein equations. We can, first, take the trace

\begin{equation}
  R +\tfrac{1}{2}(\partial\phi^{1})^{2}
  +\tfrac{1}{2}(\phi^{1})^{2}(\mathcal{D}\phi^{2})^{2} -2V
  =
  0\,,
\end{equation}

\noindent
and use it in the original equations to simplify them

\begin{subequations}
  \begin{align}
    R_{\mu\nu} +\tfrac{1}{2}\partial_{\mu}\phi^{1}\partial_{\nu}\phi^{1}
    +\tfrac{1}{2}(\phi^{1})^{2}\mathcal{D}_{\mu}\phi^{2}\mathcal{D}_{\nu}\phi^{2}
    -\tfrac{1}{2}\left(F_{\mu}{}^{\rho}F_{\nu\rho}-\tfrac{1}{4}g_{\mu\nu}F^{2}\right)
    -\tfrac{1}{2}g_{\mu\nu}V
    & =
      0\,,
  \end{align}
\end{subequations}

The components of the Ricci tensor for the above metric are

\begin{equation}
  \begin{array}{rclrcl}
    R_{tt}
    & = &
    -\tfrac{1}{2}\lambda R^{-2}\left(R^{2}\lambda'\right)'\,,\hspace{1cm}
    &
      R_{rr} & = &
                   -\lambda^{-2}R_{tt} +2R''/R\,, \\
          & & & & & \\
    R_{\theta\theta}
    & = &
          \tfrac{1}{2}\left[\lambda (R^{2})'\right]'-1
    &
      R_{\varphi\varphi}
             & = &
                   \sin^{2}{\theta}R_{\theta\theta}\,.
  \end{array}
\end{equation}

We only need to consider the $\theta\theta,tt,rr,$ components. In this order,
they are 

\begin{subequations}
  \begin{align}
  \tfrac{1}{2}\left[\lambda (R^{2})'\right]'-1
  +\tfrac{1}{4} R^{2} \left(P_{k}'\right)^{2}
  +\tfrac{1}{2}R^{2}V
  & =
    0\,,
    \\
    & \nonumber \\
    -\tfrac{1}{2}\lambda R^{-2}\left(R^{2}\lambda'\right)'
    +\tfrac{1}{2}g^{2}(\phi^{1})^{2}P_{k}^{2}
    +\tfrac{1}{4}\lambda \left(P_{k}'\right)^{2}
    -\tfrac{1}{2}\lambda V
    & =
      0\,,
    \\
    & \nonumber \\
    \tfrac{1}{2}\lambda^{-1}R^{-2}\left(R^{2}\lambda'\right)' +2R''/R
    +\tfrac{1}{2}(\phi^{1\,\prime})^{2}
    -\tfrac{1}{4}\lambda^{-1} \left(P_{k}'\right)^{2}
    +\tfrac{1}{2}\lambda^{-1}V
    & =
      0\,.
  \end{align}
\end{subequations}

Eliminating common factors etc

\begin{subequations}
  \begin{align}
  \left[\lambda (R^{2})'\right]'-2
  +\tfrac{1}{2} R^{2} \left(P_{k}'\right)^{2}
  +R^{2}V
  & =
    0\,,
    \\
    & \nonumber \\
   \left(R^{2}\lambda'\right)'
    -g^{2} R^{2}\lambda^{-1}(\phi^{1})^{2}P_{k}^{2}
    -\tfrac{1}{2} R^{2}\left(P_{k}'\right)^{2}
    + R^{2}V
    & =
      0\,,
    \\
    & \nonumber \\
    \left(R^{2}\lambda'\right)' +2RR''
    +\lambda  R^{2} (\phi^{1\,\prime})^{2}
    -\tfrac{1}{2} R^{2} \left(P_{k}'\right)^{2}
    + R^{2}V
    & =
      0\,.
  \end{align}
\end{subequations}

The difference between the last two equations is

\begin{equation}
  \label{eq:combination}
    -g^{2}\lambda^{-1}(\phi^{1})^{2}P_{k}^{2}
    -2R''/R
    -\lambda (\phi^{1\,\prime})^{2}
    =
    0\,.
\end{equation}

These equations are very difficult to solve in general. We are going to make a
simplifying assumptions that $\phi^{1}=0$ and

\begin{equation}
  \left.\partial_{\phi^{1}}V\right|_{\phi^{1}=0}
  =
  0\,,
  \,\,\,\,\,\,
  \text{and}
  \,\,\,\,\,\,
  V(\phi^{1}=0)\equiv 2\Lambda\,.
\end{equation}

Eq.~(\ref{eq:phi1eq-2}) is solved automatically and the combination
Eq.~(\ref{eq:combination}) is solved by

\begin{equation}
R = ar\,,  
\end{equation}

\noindent
where we have eliminated an integration constant through a shift of $r$ and
where the integration constant $a$ will be set to 1 to give metric at spatial
infinity the standard normalization.  We are left with the following
equations:

\begin{subequations}
  \begin{align}
  \left(r^{2}P_{k}'\right)'
  & =
    0\,,
    \\
    & \nonumber \\
  2\left(\lambda r \right)'-2
  +\tfrac{1}{2} r^{2} \left(P_{k}'\right)^{2}
  +2\Lambda r^{2}
  & =
    0\,,
    \\
    & \nonumber \\
    \left(r^{2}\lambda'\right)' 
    -\tfrac{1}{2} r^{2}\left(P_{k}'\right)^{2}
    +2\Lambda r^{2}
    & =
      0\,.
  \end{align}
\end{subequations}

The first equation is solved by

\begin{equation}
  \label{eq:Pkprime}
  P_{k}'
  =
  \frac{a}{r^{2}}\,,
\end{equation}

\noindent
for some other integration constant that we call, again, $a$. Then the other
two equations take the form

\begin{subequations}
  \begin{align}
    2\left(\lambda r \right)'-2
  +\tfrac{1}{2} a^{2}r^{-2} 
  +2\Lambda r^{2}
  & =
    0\,,
    \\
    & \nonumber \\
    \left(r^{2}\lambda'\right)' 
    -\tfrac{1}{2} a^{2}r^{-2}
    +2\Lambda r^{2}
    & =
      0\,.
  \end{align}
\end{subequations}

Combining these two equations we can eliminate the terms that depend on $a$:

\begin{equation}
  \left(r^{2}\lambda\right)'' -2 +4\Lambda r^{2}
  =
  0\,.
\end{equation}

We can integrate it immediately:

\begin{equation}
  \lambda= 1 +\frac{b}{r}+\frac{c}{r^{2}}-\frac{\Lambda}{3} r^{2}\,,
\end{equation}

\noindent
which corresponds to the Reissner-Nordstr\"om-(anti-)De Sitter (RN(A)DS)
metric \cite{Tangherlini:1963bw}. As a matter of fact, substituting the above
value of $\lambda$ in either of the previous equations, we find that

\begin{equation}
  c
  =
  a^{2}/4\,,
\end{equation}

\noindent
and, since $a$ is, up to constants, the electric charge, the identification of
the solution with the RN(A)DS solution is confirmed.


\end{document}